\documentclass[a4paper,10pt,aps,prd,nofootinbib,superscriptaddress]{revtex4-2}
\usepackage[margin=0.9in]{geometry}

\usepackage{slashed}
\usepackage{caption}
\usepackage{subfigure}
\usepackage{amsmath,amssymb,graphicx,xcolor,mathtools}
\usepackage[colorlinks=true,linktocpage=green,linkcolor=blue,citecolor=blue]{hyperref}
\numberwithin{equation}{section}
\allowdisplaybreaks
\usepackage{hyperref}
\hypersetup{
	colorlinks=true,
	linkcolor=blue,
	filecolor=magenta,      
	urlcolor=cyan,
	pdftitle={HQB-M}
}

\usepackage{mathrsfs}

\def\be {\begin{equation}}
	\def\ee {\end{equation}}
\def\bea {\begin{eqnarray}}
	\def\eea {\end{eqnarray}}
\def\bc {\begin{center}}
	\def\ec {\end{center}}
\def\nn {\nonumber}

\def\om {\omega}

\DeclareMathAlphabet{\mathpzc}{OT1}{pzc}{m}{it}

\begin{document}
	
	\title{Spatial diffusion of heavy quarks in background magnetic field}
	
	\author{Sarthak Satapathy}

	\author{Sudipan De}
	\email{Corresponding Author Email: sudipan86@gmail.com}
	\affiliation{Department of Physics, Dinabandhu Mahavidyalaya, Bongaon, North 24 Parganas - 743235, West Bengal, India
	}
	\author{Jayanta Dey}
	\affiliation{Department of Physics, Indian Institute of Technology Indore, Simrol, Indore 453552, India
	}
	
	\author{Sabyasachi Ghosh}
	\affiliation{Indian Institute of Technology Bhilai, GEC Campus \\
		Sejbahar, Raipur - 492015, Chattisgarh, India
	}

	


	\begin{abstract}
The ratio of shear viscosity to entropy density shows a valley-shaped pattern well-known in the community of heavy-ion physics. Diffusion coefficients of heavy quark and meson shows the similar structure, and both sketches have become quite popular in the community. Present work has attempted a finite magnetic field extension of the diffusion
coefficients of heavy quark and meson. Using Einstein's diffusion relation, we calculated heavy quark and heavy meson diffusion by the ratio of conductivity to susceptibility in the kinetic theory framework of relaxation time approximation. The relaxation time of heavy quark and meson
are tuned from the knowledge of earlier works on spatial diffusion  estimations, and then we have extended the framework for a finite magnetic field,
where our outcomes have revealed two aspects - anisotropic and quantum aspects of diffusion with
future possibilities of phenomenological signature.
		
	\end{abstract}
	
	\maketitle

	\section{Introduction}
	\label{INTRO}
	During collision of two nuclei at relativistic energies, it is expected that a hot and dense state of matter known as Quark Gluon Plasma (QGP)~\cite{Shuryak:2004cy} is formed. Heavy quarks, namely charm ($c$) and bottom $(b)$ quarks are considered one of the fine probes of QGP due to the the fact that their masses $(M)$ are significantly larger than the QCD energy scale ($\Lambda_\text{QCD}$) where $\Lambda_\text{QCD} \approx 200$ MeV and the temperature $T$ at which QGP is created. Unlike the light quarks they do not thermalize quickly and witness the entire evolution of the fireball. One of the most important observable to study the QGP is transverse momentum suppression ($R_{AA}$) of heavy quarks which lead us to the drag and diffusion coefficient~\cite{Prino:2016cni,Rapp:2018qla,Das:2010tj}. Currently it is estimated that a very strong magnetic field is created at very early stage of heavy-ion collisions~\cite{Skokov:2009qp,Bzdak:2011yy} where the estimated values of the magnetic field created at RHIC and LHC is of the order of $10^{18}$ to $10^{19}$ Gauss~\cite{Tuchin:2013ie}. Strong magnetic fields are known to influence relativistic fluids when $eB \gg \Lambda_\text{QCD}^2$ and some pronounced effects connected with QGP phenomenology, such as flow~\cite{Das:2016cwd}, chiral magnetic effect~\cite{Kharzeev:2007jp}, jet quenching coefficient $\hat{q}$~\cite{Banerjee:2021sjm}, diffusion coefficients of charm quarks\cite{Fukushima:2015wck,Finazzo:2016mhm,Goswami:2022szb}, etc have been observed. Present work  focusses on the diffusion phenomenology of heavy quark and meson into QGP and hadronic matter at finite magnetic field.

	The influence of magnetic field on charmonium studied through holographic QCD~\cite{Dudal:2014jfa}, is one of the initial attempts to study the dynamics of heavy quarks in an anisotropic medium, where the influence of background magnetic field on the melting of $J/\psi$ has been studied. As an extension of this work the transport properties of $J/\psi$ vector mesons and heavy quarks, particularly spatial diffusion and quark number susceptibility have been studied as a function of temperature and magnetic field in the framework of holographic QCD model in Ref.~\cite{Dudal:2018rki} by the help of soft wall model which is a simpler version of holographic QCD. Their results show that spatial diffusion splits into two components viz. longitudinal and transverse relative to the direction of magnetic field. 
	The origin of anisotropy in Ref.~\cite{Dudal:2018rki} is due to magnetic field which does not affect the longitudinal component but affects the transverse component of spatial diffusion. Previously these anisotropic factors carrying a magnetic field dependence have been found to be present in transport coefficients of QGP in background magnetic 
field~\cite{Dey:2019axu,Satapathy:2021cjp,Dey:2020awu,Dash:2020vxk,Nam:2012sg,Hattori:2016cnt,Hattori:2016lqx,Harutyunyan:2016rxm,Kerbikov:2014ofa,Feng:2017tsh,Fukushima:2017lvb,Li:2018ufq,Das:2019wjg,Das:2019ppb,Ghosh:2019ubc,Satapathy:2021wex,Ghosh:2020wqx,Li:2017tgi,Nam:2013fpa,Alford:2014doa,Tawfik:2016ihn,Tuchin:2011jw,Ghosh:2018cxb,Mohanty:2018eja,Dey:2019vkn,Hattori:2017qih,Huang:2009ue,Huang:2011dc,Agasian:2011st,Agasian:2013wta} 
	particularly in  conductivity~\cite{Dey:2019axu,Satapathy:2021cjp,Dey:2020awu,Dash:2020vxk,Nam:2012sg,Hattori:2016cnt,Hattori:2016lqx,Harutyunyan:2016rxm,Kerbikov:2014ofa,Feng:2017tsh,Fukushima:2017lvb,Li:2018ufq,Das:2019wjg,Das:2019ppb,Ghosh:2019ubc,Satapathy:2021wex}. In order to understand this connection with the results of holographic QCD it is to be noted that diffusion in condensed matter physics can be expressed as a ratio of  conductivity and susceptibility~\cite{Romatschke:2017ejr}, whose connection for the relativistic case of QGP can be made through Kubo formula~\cite{Laine:2016hma}. In this work we have employed this idea to understand the behaviour of spatial diffusion in magnetic field. 
	In the presence of magnetic field, the anisotropic factors appearing in conductivity match with those of diffusion, which have been obtained via holographic QCD in Ref.~\cite{Dudal:2018rki}.  These anisotropic factors govern the diffusive dynamics of charm quarks and $D^+$ mesons in a magnetized medium with their temperature and magnetic field dependence. 
	
	
	In the absence of magnetic field, we can find a long list of Refs.~\cite{Berrehrah:2014tva,Riek:2010fk,vanHees:2007me,Liu:2016ysz,Scardina:2017ipo,Banerjee:2011ra} which have 
	estimated spatial and momentum diffusion of heavy quark through QGP via different methodologies
	like Dynamical Quasi-Particle Model (DQPM)~\cite{Berrehrah:2014tva}, $T$-matrix approach~\cite{Riek:2010fk,vanHees:2007me,Liu:2016ysz}, quasi-particle model (QPM)~\cite{Scardina:2017ipo}, 
	Lattice Quantum Chromodynamics (LQCD)~\cite{Banerjee:2011ra}. 
	Along with the QGP phase, spatial and momentum diffusion of $D$~\cite{Ghosh:2011bw,Tolos:2013kva,He:2011yi,Torres-Rincon:2021yga,Abreu:2011ic} meson, $B$~\cite{Das:2011vba,Abreu:2012et} meson and $\Lambda_c$~\cite{Ghosh:2014oia,Tolos:2016slr} baryon, which lie in the 
	hadronic phase, have also be studied. Collection of hadronic~\cite{Ghosh:2011bw,Tolos:2013kva,He:2011yi,Torres-Rincon:2021yga,Abreu:2011ic}
	and quark~\cite{Berrehrah:2014tva,Rapp:2008qc,vanHees:2007me,Liu:2016ysz,Scardina:2017ipo,Banerjee:2011ra} temperature domain estimations of normalized spatial diffusion for $D^+$ meson have unfolded an interesting u-shape or valley-shape pattern. It was first pointed out in 2011 by Ref.~\cite{Abreu:2011ic} and later
	it became well popular in community~\cite{Torres-Rincon:2021yga,Prino:2016cni,Rapp:2018qla}. Similar pattern was first pointed in 2006 by Ref.~\cite{Csernai:2006zz} for normalized shear viscosity or shear viscosity to entropy density ratio by combining pQCD and ChPT results. Later this pattern was verified by different effective QCD model calculations~\cite{Abhishek:2017pkp,Singha:2017jmq,Sasaki:2008um,Deb:2016myz,Chakraborty:2010fr}. A finite magnetic
	field extension for normalized shear viscosity or shear viscosity to entropy density ratio has been explored by recent Refs.~\cite{Finazzo:2016mhm,Dash:2020vxk,Dey:2019axu,Das:2019ppb,Tuchin:2011jw,Ghosh:2018cxb,Mohanty:2018eja,Dey:2019vkn,Ghosh:2020wqx},
	which are probably seeking further research (by using alternative model estimations) for getting converged conclusion. In this regard, the finite magnetic field extension of normalized spatial diffusion for 
	$D^+$ meson has not been explored explicitly. Only Ref.~\cite{Dudal:2018rki} has provided a holographic estimation of heavy quark diffusion at finite magnetic field and Ref.~\cite{Fukushima:2015wck}
	have provided its leading order QCD estimation and pointed out its possible phenomenological impact. Present work is aimed to explore the finite magnetic field extension of u-shape or valley-shape 
	pattern of normalized spatial diffusion. With the help of the kinetic theory framework of heavy quark conductivity and diffusion, we have first attempted to calibrate our estimation with the existing
	Refs.~\cite{Berrehrah:2014tva,Rapp:2008qc,vanHees:2007me,Liu:2016ysz,Scardina:2017ipo,Banerjee:2011ra} for quark temperature and Refs.~\cite{Ghosh:2011bw,Tolos:2013kva,He:2011yi,Torres-Rincon:2021yga,Abreu:2011ic} for hadronic temperature domains by tuning
	charm quark and $D^+$ meson relaxation time. Then we have done its finite magnetic field extension, where cyclotron time scale will add with relaxation time via Boltzman equation and we 
	get an effective relaxation time.


	This paper is organized as follows. In Sec.~\ref{B=0} we provide Kubo formula and the relaxation time approximation expressions of  conductivity, susceptibility and spatial diffusion in the absence of magnetic field. In Sec.~\ref{formalism:f} we have studied spatial diffusion in the presence of background magnetic field. In Sec.~\ref{sec:results} we have numerically analysed the results of our work for the isotropic case ($B=0$) and anisotropic case ($B\neq 0$) as a function of temperature and magnetic field for $D^+$ mesons and charm quarks. Finally summary and conclusions with an outlook for future research has been presented in Sec.~\ref{summary}. In the end we provide the details of the calculation of  conductivity in magnetic field, spatial diffusion and Kubo formula for spatial diffusion in Appendix~\ref{sec:ECB} and Appendix~\ref{QNS}.   \\
	
	Throughout the paper we have used the metric convention $g^{\mu\nu} = \text{diag}(1,-1,-1,-1)$ and $\hbar, k_\text{B}, c = 1$.

	\section{Formalism}	
	\label{Form} 
	In this section we will quickly address the formalism for spatial diffusion, whose detailed calculations are given in appendix.
	The section is further divided into two subsections for the framework in absence and presence of magnetic field respectively.
	\subsection{Heavy quark diffusion at $B = 0$}
	\label{B=0}
	We know that spatial diffusion $D$ of charm quark and its corresponding conductivity $\sigma$ and susceptibility $\chi$ are inter-connected as~\cite{Romatschke:2017ejr,Teaney}
	\bea
	D = \frac{\sigma}{\chi}~.
	\label{kubo-0}
	\eea 
	For quantum field theoretical structure~\cite{Laine:2016hma} the spatial diffusion coefficient can be defined through Kubo relation  
	\bea
	D = \frac{1}{3\chi}\displaystyle{\lim_{q_0\to 0^+}}\frac{\rho^{ii}(q_0,\vec{0})}{q_0}~,
	\label{kubo-1}
	\eea 
	where 
	$$\rho^{ii}(q_0,\vec{q}) = \text{Im~}i\int d^4x~ e^{iq\cdot x}\big\langle J^i(x) J^i(0) \big\rangle_{\beta}$$ 
	is the two-point spectral function of heavy quark currents $J^i(x)$, $q_0$ is the energy and $\big\langle\mathcal{\hat{O}}\big\rangle_\beta$ is the ensemble average of operators (say $\mathcal{\hat{O}}$) in thermal field theory. In accordance with Kubo formalism, Eq.~\ref{kubo-1} translates to the Kubo formula for diffusion which is given by Eq.~(\ref{kubo-0}), 
	where we have substituted the expression, $\sigma = \frac{1}{3}\displaystyle{\lim_{\omega\to 0^+}}\frac{\rho^{ii}(\omega,\vec{0})}{\omega}$, where $\sigma$ is the heavy quark conductivity.   
	For the calculation of heavy quark diffusion coefficient one requires the expressions of heavy quark conductivity and heavy quark susceptibility, which can be calculated from two formalisms viz. Kubo formalism and RTA formalism. 
	In the absence of background magnetic field, the expression of heavy quark conductivity (any other charge conductivity) obtained via Kubo and RTA formalism is the same, while in the presence of magnetic field they can be different, e.g. for  conductivity, see Refs.~\cite{Satapathy:2021cjp,Satapathy:2021wex}. The expression of heavy quark conductivity (see Appendix~\ref{sec:ECB}) in the absence of magnetic field obtained from RTA is given by  
	\bea
	\sigma = \frac{g\beta}{3}\int\frac{d^3k}{(2\pi)^3}\frac{\vec{k}^2}{\omega_k^2}\tau_c~ f_0\big(1 - f_0\big)
	\label{sigma_B0}
	\eea 
	and the expression of susceptibility (see Appendix~\ref{QNS}) is given by 
	\bea
	\chi = g\beta\int\frac{d^3k}{(2\pi)^3}f_0(1-f_0)
	\label{chi_B0}
	\eea 
	where $\beta = T^{-1}$  and $T$ is the temperature, $\omega = \sqrt{\vec{k}^2 + m^2}$ is the energy, $m$ is the heavy quark mass, $f_0$ is the Fermi-Dirac (FD) distribution function given by $f_0 = \big[e^{\beta(\omega-\mu_c)}+1\big]^{-1} $, $\mu_c$ is the charm chemical potential, $\tau_c$ is the heavy quark relaxation time and $g$ is charm quark degeneracy factor. Below quark-hadron transition temperature, we have to consider heavy meson i.e., $D^+$ in place of heavy 
	quark. So, in Eqs~(\ref{sigma_B0}) and (\ref{chi_B0}), mass of $c$ quark will be replaced by $D^+$ meson mass, degeneracy factor $g=6$ of $c$ quark will be replaced by 
	degeneracy factor $g=1$ of $D^+$ meson and FD distribution will be modified to Bose-Einstein (BE) distribution function. Hence, for $D^+$ meson, we have to use the expressions:
	\bea
	\sigma = \frac{\beta}{3}\int\frac{d^3k}{(2\pi)^3}\frac{\vec{k}^2}{\omega_k^2}\tau_c~ f_0\big(1 + f_0\big)
	\label{sigma_B0_D}
	\eea 
	and
	\bea
	\chi = \beta\int\frac{d^3k}{(2\pi)^3}f_0(1 + f_0)~.
	\label{chi_B0_D}
	\eea 
	\subsection{Heavy quark diffusion at $B \neq 0$}
	\label{formalism:f}
	Let us consider a background magnetic field $\vec{B} = B~\hat{k}$ pointing in the $z$-direction. In absence of magnetic field, we can express any conductivity tensor as $\sigma_{ij} = \sigma \delta_{ij}$, while in presence of magnetic field, we can get an anisotropic  conductivity tensor~\cite{Dey:2020awu,Harutyunyan:2016rxm} (see Appendix~\ref{sec:ECB})
	\be 
	\sigma_{ij}=\sigma_0 ~\delta_{i j} - \sigma_1 ~\epsilon_{ijk} b_k + \sigma_2 ~b_i b_j~,
	\ee 
	where $b_i$ is a unit vector along the magnetic field, $\epsilon_{ijk}$ is the Levi-Civita symbol and spatial component of all Lorentz indices $(i, j, k)$ can vary as $x,y,z$. The non-zero components of anisotropic conductivity tensor have inter-connected relation and can be identified as~\cite{Dey:2020awu,Harutyunyan:2016rxm}  
	\bea 
	&&{\rm perpendicular/Transverse~ component:}~ \sigma_{xx}=\sigma_{yy}=\sigma_{0}=\sigma_{\perp}~,
	\nn\\
	&&{\rm Hall~ component:}~\sigma_{xy} = -\sigma_{yx}=\sigma_{1}=\sigma_{\times}~,
	\nn\\
	&&{\rm parallel/Longitudinal~ component:}~\sigma_{zz}=\sigma_{0}+\sigma_{2}=\sigma_{\parallel}~.
	\eea 
	According to Einstein's relation~\cite{Romatschke:2017ejr}, the spatial diffusion coefficient $(D_{ij})$ can be expressed as a ratio of  conductivity $(\sigma_{ij})$ and susceptibilty $(\chi)$ of the medium. These quantities become anisotropic in the presence of magnetic field thus taking a $3\times 3$ matrix structure given by  
	\bea
	D_{ij} = \frac{\sigma_{ij}}{\chi}~.
	\eea 
	Following RTA approach the longitudinal and transverse components of conductivity for $c$ quark and $D^+$ meson are given by~\cite{Dey:2020awu,Harutyunyan:2016rxm} (see Appendix~\ref{sec:ECB})
	\bea
	&&\sigma_{zz} = \sigma_\parallel = \frac{g\beta}{3}\int\frac{d^3k}{(2\pi)^3}\frac{k^2}{\omega_k^2}\tau_cf_0\big[1\mp f_0\big] 
	\label{f:RTAsigmazz} \\
	&&\sigma_{xx} = \sigma_{yy} =\sigma_\perp = \frac{g\beta}{3}\int\frac{d^3k}{(2\pi)^3}\frac{k^2}{\omega_k^2}\frac{\tau_c}{ 1 + \frac{\tau_c^2}{\tau_B^2}}f_0\big[1\mp f_0\big],
	\label{f:RTAsigmaxx}
	\eea
	and the susceptibility are given by 
	\bea
	\chi = g\beta\int\frac{d^3k}{(2\pi)^3}f_0(1\mp f_0)~,
	\label{f:susc}
	\eea 
	where $g=6$, FD distribution will be taken for $c$ quark and $g=1$, BE distribution will be taken for $D^+$ meson
	Here, $\tau_{B}=\frac{\omega}{q B}$ with $q=\frac{2e}{3}$ for $c$ quark and $q=e$ for $D^+$ meson are their respective inverse of cyclotron frequency. Here, $\sigma_{zz}$ (or $\sigma_\parallel$) is the longitudinal component of  conductivity i.e., parallel to the magnetic field and $\sigma_{xx}$ (or $\sigma_\perp$) is the transverse component of  conductivity which have been derived via RTA in Appendix~\ref{sec:ECB}. The absence of Landau quantization of energies in RTA expressions prompts us to call them as classical results. On the other hand, on applying Landau quantization of energies and quantizing the phase space part of the momentum integral which has been shown in Appendix~\ref{sec:ECB}, the expressions of Eqs.~(\ref{f:RTAsigmazz}), (\ref{f:RTAsigmazz}) and (\ref{f:susc}) for $c$ quark are given by
	\bea
	\sigma_\perp^\text{QM} &=& \frac{3}{T}\sum_{l=0}^\infty (2-\delta_{l,0})\frac{qB}{2\pi}\int_{-\infty}^{+\infty}\frac{dk_z}{2\pi}\frac{lqB}{\omega_l^2}\tau^\perp f_0(1-f_0)
	\label{f:ecqm:perp}\\
	\sigma_\parallel^\text{QM} &=& \frac{3}{T}\sum_{l=0}^\infty (2-\delta_{l,0})\frac{qB}{2\pi}\int_{-\infty}^{+\infty}\frac{dk_z}{2\pi}\frac{k_z^2}{\omega_l^2} \tau^\parallel f_0(1-f_0),
	\label{f:ecqm:pll}
	\eea 
	\bea
	\chi_\text{QM} = 3\beta\sum_{l=0}^{\infty}(2-\delta_{l,0})\frac{qB}{2\pi}\int_{-\infty}^{+\infty}\frac{dk_z}{2\pi}~f_0(1-f_0)~,
	\label{f:susc:qm}
	\eea  
	where the superscript QM denotes the calculations or expressions where Landau quantization is taken into account, $l$ is the Landau level index, $\tau^\parallel = \tau_c$ is the relaxation time for the longitudinal component, $\tau^\perp = \frac{\tau_c}{1 + \frac{\tau_c^2}{\tau_B^2}}$ is the relaxation time for the transverse component, $\tau_c$ is the relaxation time in the absence of magnetic field and $\omega_l = \sqrt{k_z^2 + m^2 + 2lqB}$ is the Landau quantized energy with $q=\frac{2}{3}e$ for $c$ quark.
	
	Corresponding expressions for $D^+$ meson are given by
	\bea
	\sigma_\perp^\text{QM} &=& \frac{1}{T}\sum_{l=0}^\infty \frac{qB}{2\pi}\int_{-\infty}^{+\infty}\frac{dk_z}{2\pi}\frac{(l+1/2)qB}{\omega_l^2}\tau^\perp f_0(1 +f_0)
	\label{f:ecqm:perp}\\
	\sigma_\parallel^\text{QM} &=& \frac{1}{T}\sum_{l=0}^\infty \frac{qB}{2\pi}\int_{-\infty}^{+\infty}\frac{dk_z}{2\pi}\frac{k_z^2}{\omega_l^2} \tau^\parallel f_0(1+f_0),
	\label{f:ecqm:pllD}
	\eea 
	\bea
	\chi_\text{QM} = \beta\sum_{l=0}^{\infty}\frac{qB}{2\pi}\int_{-\infty}^{+\infty}\frac{dk_z}{2\pi}~f_0(1+f_0)~,
	\label{f:susc:qmD}
	\eea
	where $\omega_l = \sqrt{k_z^2 + m^2 + (2l+1)qB}$ is the Landau quantized energy with $q=e$ for $D^+$ meson.

	\section{Results and Discussions}
	\label{sec:results}
	
	In this section we have investigated the results of our work on spatial diffusion in the presence and absence of background magnetic field for exploring the impact of magnetic field on spatial diffusion of heavy quark and meson. 
	The results for the diffusion have been calculated from the ratio of conductivity to susceptibility in the RTA and QM formalism. The results of RTA in Eqs.~(\ref{f:RTAsigmazz}), (\ref{f:RTAsigmaxx}) and (\ref{f:susc}) do not carry the information of Landau quantization, whereas the quantum theoretical results of Eqs.~(\ref{f:ecqm:pll}), (\ref{f:ecqm:perp}) and (\ref{f:susc:qm}) carry the contribution of all Landau levels.

	The similarity of QM results with RTA results is reflected in the structure of the anisotropic factors which carry the information of cyclotron frequency ($\tau_B^{-1} = \frac{\omega}{eB}$), where $\omega$ is quantized for QM results. For temperatures between the temperature range 0.2 GeV to 0.4 GeV charm quarks are the dominant particles contributing to diffusion, whereas between 0.1 GeV to 0.18 GeV, $D^+$ mesons are the particles contributing to diffusion.
	
	
	%
	\begin{figure}[h]
		\includegraphics[scale = 0.32]{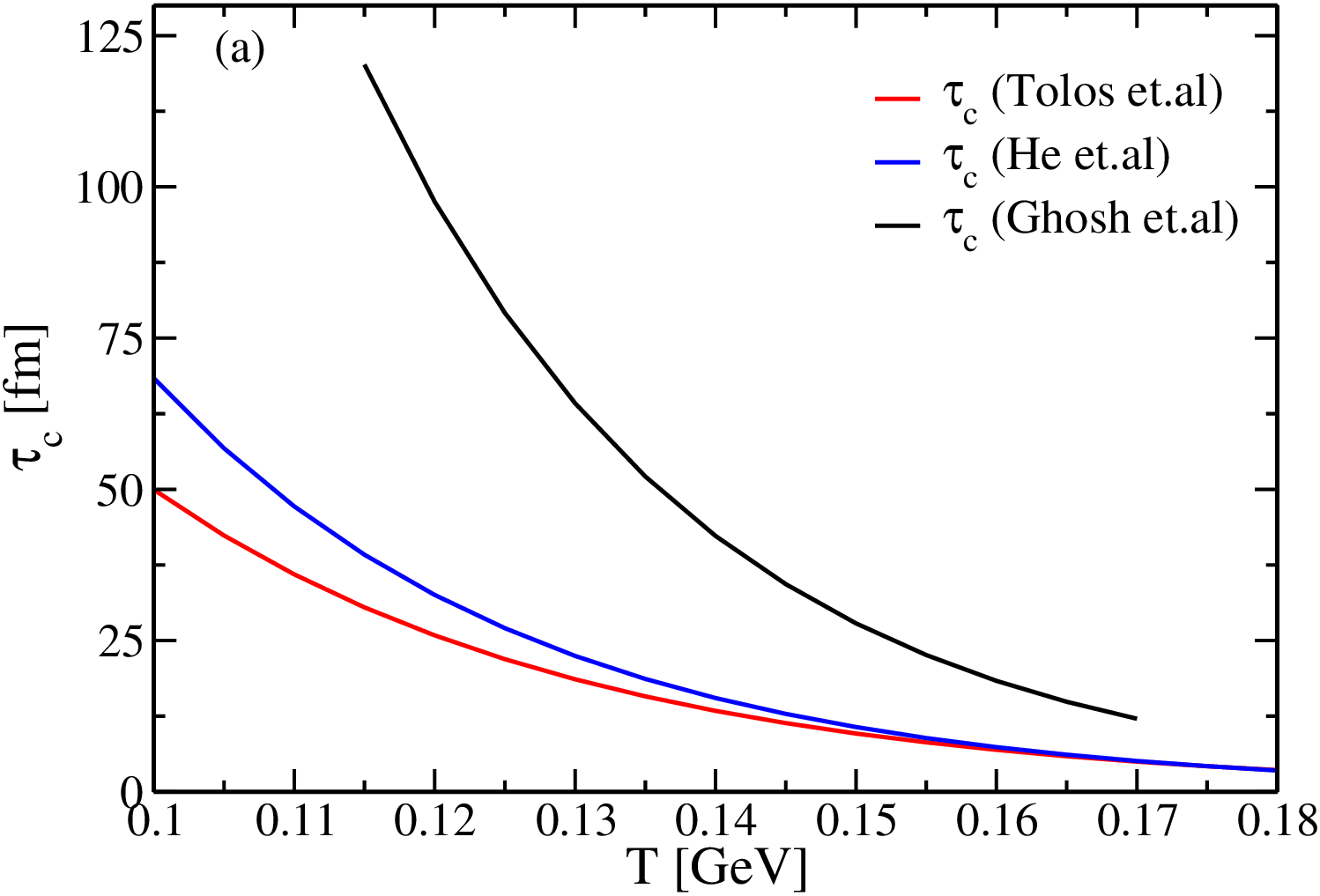}
		\includegraphics[scale = 0.32]{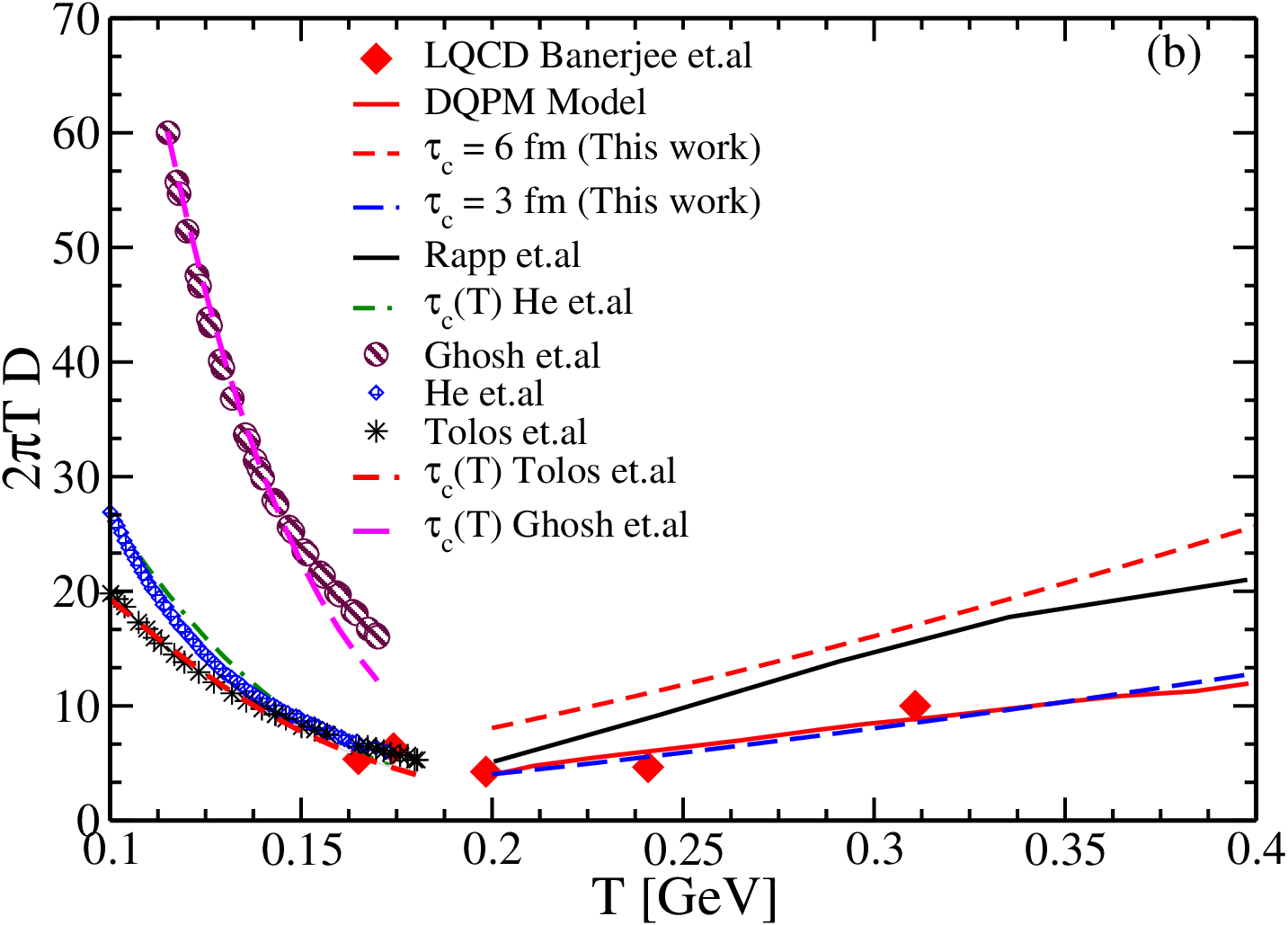}
		\caption{(a)Relaxation time parameterized as $\tau_c(T)$ by fitting the results, obtained by Ghosh et al.~\cite{Ghosh:2011bw} (Black solid line), He et al.~\cite{He:2011yi} (blue solid line) and Tolos et al.~\cite{Tolos:2013kva} (red solid line). (b) Their estimated spatial diffusion coefficient ($D$) for $D^+$ mesons and charm quark diffusion coefficient, obtained by Refs.~\cite{Banerjee:2011ra,Berrehrah:2014tva,Rapp:2008qc}, which can be mapped by tuning $\tau_c=3$-$6$ fm.}
		\label{Models-tauc}
	\end{figure}
    Since our focal interest to see the modification in heavy quark/meson conductivity and diffusion due to finite magnetic field,
    so we will first tune our results with the existing results for $B = 0$, then we will see their modification based on their finite magnetic field expressions. We have plotted the dimensionless quantity $2\pi TD$ vs $T$ in Fig.~\ref{Models-tauc}(b), where $2\pi T$ stands for the thermal wavelength. 
	%
	In Fig.~\ref{Models-tauc}(b) for hadronic temperature domain, we have included the data points, obtained by Ghosh~{\it et.al}~\cite{Ghosh:2011bw} (circles), He {\it et.al}~\cite{He:2011yi} (blue diamonds), Tolos {\it et.al}~\cite{Tolos:2013kva} (stars).
	Using RTA Eqs.~(\ref{kubo-0}), (\ref{sigma_B0}), (\ref{chi_B0}), we have fitted $2\pi TD$ of Refs.~\cite{Ghosh:2011bw, He:2011yi, Tolos:2013kva} by tuning $\tau_c(T)$, which has been shown in Fig.~\ref{Models-tauc}(a). 
	%
	Then for the quark temperature, we have included LQCD data of Banerjee et al.~\cite{Banerjee:2011ra} (red diamonds), DQPM estimation of Berrehra et. al.~\cite{Berrehrah:2014tva} (red solid line) and potential model of Rapp et al.~\cite{Rapp:2008qc} (black solid line) in  
	Fig.~\ref{Models-tauc}(b). Reader may find many more alternative model estimation in quark temperature domain but our aim is just to get
	a rough ranges of spatial diffusion coefficients and tune our heavy quark relaxation time accordingly.
	These all non-pQCD model estimations are quite smaller than pQCD values~\cite{Moore:2004tg} and in favor of experimental data~\cite{Prino:2016cni,Rapp:2018qla}.
	By considering constant $\tau_c=3-6$ fm, we can crudely cover the non-pQCD temperature domain $T=0.2$-$0.4$ GeV.
	%
	
	After tuning the order of magnitude and qualitative $T$-dependence of spatial diffusion, we  proceed for their finite magnetic field extension, which is the main aim of this work. 
	We should note that spatial diffusion is the ratio of heavy quark/meson conductivity and susceptibility, where conductivity is affected in the presence of background magnetic field taking a multi-component structure but susceptibility remains unaffected. Anisotropy in conductivity is induced through the relaxation times in  conductivities and spatial diffusions i.e., the longitudinal/parallel relaxation time $\tau_c^\parallel = \tau_c$ and the transverse/perpendicular relaxation time $\tau_c^\perp = \tau_c\Big/\Big( 1 + \frac{\tau_c^2 e^2B^2}{\omega^2}\Big)$, where $\tau_c$ is the relaxation time in the absence of magnetic field. A similar anisotropic factor is present in the spatial diffusion in magnetic field which has been obtained in AdS/CFT~\cite{Dudal:2018rki} given by 
	$$D_\parallel = \frac{T}{\gamma m}~~ \text{and}~~ D_\perp = \frac{D_\parallel}{1 + \frac{q^2B^2}{m^2\gamma^2}}$$ where $D_\parallel$ and $D_\perp$ are the longitudinal and transverse components of spatial diffusion, $q$ is the charge, $m$ is the mass and $\gamma$ plays the role of $\tau_c^{-1}$ of heavy quarks in the AdS/CFT framework. The origin of these anisotropic factors is connected to the action of Lorentz force on the charged particles along the longitudinal and transverse directions of magnetic field. The detailed plots and discussions on these two components are discussed one by one in next two subsections.

	
	\subsection{Longitudinal components}	
	In this subsection, we will discuss about the longitudinal/parallel component of conductivity first and then diffusion by using their RTA and QM expressions. In between the RTA and QM results of conductivity and diffusion, results of heavy quark/meson susceptibility is also discussed. 
	
\begin{figure}[h]
		\includegraphics[scale = 0.35]{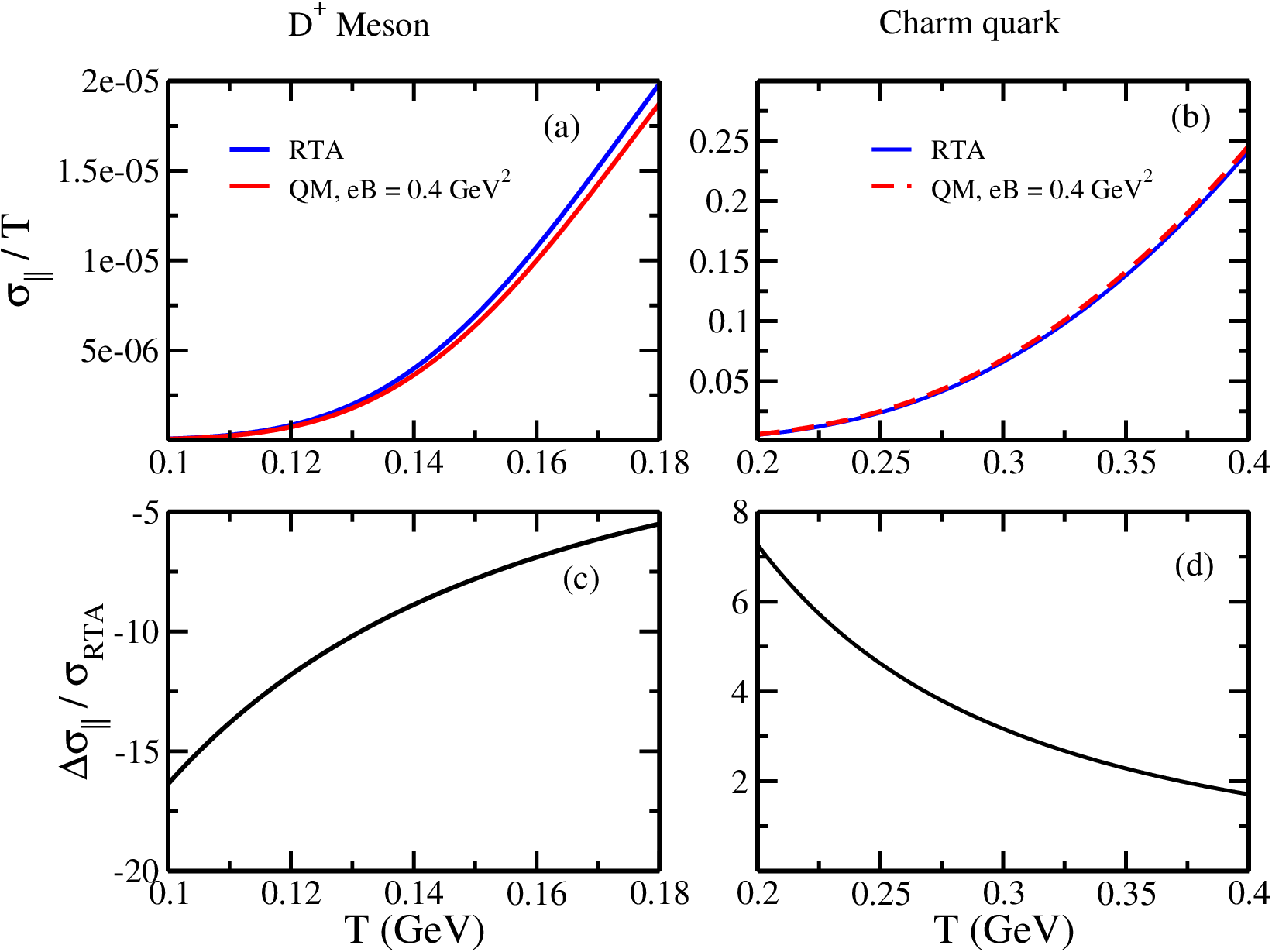}
		\caption{Normalized values of longitudinal conductivity $\sigma_{\parallel} / T$ plotted as a function of $T$ for RTA and QM results for (a) $D^+$ mesons and (b) charm quarks at $eB = 0.4$ GeV$^2$. Relative percentage changes of $\sigma_{\parallel}$ of QM results with respect to RTA results for (c) $D^+$ mesons and (d) charm quarks at $eB = 0.4$ GeV$^2$.  }
		\label{ec:pll-T}
	\end{figure}

\begin{figure}[h]
		\includegraphics[scale = 0.35]{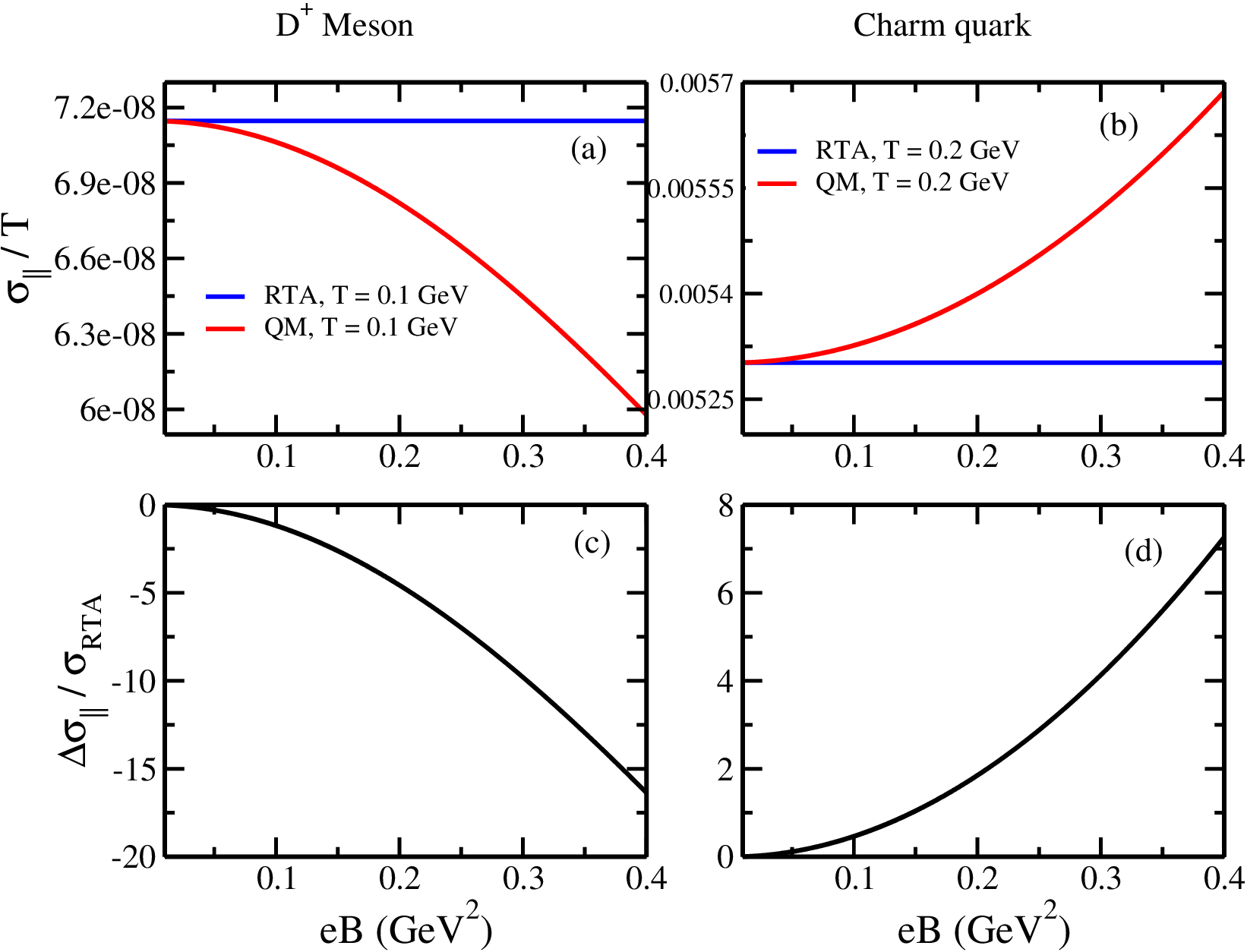}
		\caption{ Normalized values of longitudinal conductivity $\sigma_{\parallel} / T$ plotted as a function of $eB$ for RTA and QM results for (a) $D^+$ mesons at $T = 0.1$ GeV and (b) charm quarks at $T = 0.2$ GeV. Relative percentage changes of $\sigma_{\parallel}$ of QM results with respect to RTA results for (c) $D^+$ mesons at $T = 0.1$ GeV (d) charm quarks at $T = 0.2$ GeV. }
		\label{ec:pll-B}
	\end{figure}

	Let us first discuss $\sigma_{\parallel} / T$ vs $T$, $eB$ plots, given in Figs.~\ref{ec:pll-T} and \ref{ec:pll-B} respectively. For parallel component, RTA expression (\ref{f:ecqm:pll}) in presence of magnetic field is exactly the same with zero magnetic field expression in Eq.(\ref{sigma_B0}).
	In Fig.~\ref{ec:pll-T}(a) and (b) we have plotted $\sigma_{\parallel} / T$ vs $T$ for $D^+$ mesons and charm quarks at $eB = 0.4$ GeV$^2$. 
	For the results of the hadronic zone we have taken the parameterized relaxation time of the results of He {\it et.al}~\cite{He:2011yi} $\tau_c(T)_\text{Rapp} = 2794.15 ~e^{-37~T}$~ fm and for quark phase, we have considered constant relaxation time $\tau_c=6$ fm, which approximately fits data of Rapp et al.~\cite{Rapp:2008qc}.
	%
	%
	Now we focus on the results of classical (RTA) and quantum theoretical (QM) expressions in Fig.~\ref{ec:pll-T}(a) and (b) where the RTA results are shown by the blue solid curves and the QM results are shown by the red curves(solid for $D^+$ meson and dashed for $c$ quark). 
 In the lower panel of Fig.~\ref{ec:pll-T} i.e., in Fig.~\ref{ec:pll-T}(c) and (d) we have plotted $\Delta\sigma / \sigma_\text{RTA}$ where $\Delta\sigma = (\sigma_\text{QM} - \sigma_\text{RTA})\times 100$ which qualitatively shows the relative change of  conductivity for the QM case with respect to the RTA case. 
 The QM results become different from RTA results due to Landau quantization and we notice the quantum enhancement for charm quark (Fermion)
 and quantum suppression for $D^+$ meson (Boson), which are decreasing with temperature. The facts may be connected with 
 the Landau level sum of BE/FD distribution functions $f_0=1/\Big[e^{\beta\omega_l}\pm 1\Big]$ with their corresponding quantized energy $\omega_l$, which ultimately give the non-trivial results of suppression/enhancement for $D^+$ meson (Boson)/charm quark (Fermion) due to their corresponding Landau quantization effect.
%
	%

	To understand the behaviour of  conductivity with magnetic field we plotted $\sigma_{\parallel} / T$ with $eB$ in Fig.~\ref{ec:pll-B}. In the upper panel we have shown Fig.~\ref{ec:pll-B}(a) and (b) where the variation of $\sigma_{\parallel} / T$ with $eB$ has been shown for $D^+$ mesons and charm quarks on the left and right panel for $eB$ between 0.01 GeV$^2$ to 0.4 GeV$^2$. 
	The results for $D^+$ mesons have been shown in Fig.~\ref{ec:pll-B}(a) where the blue solid line is the RTA result and the red solid line is the QM result. The results for charm quarks have been shown in Fig.~\ref{ec:pll-B}(b) where the blue solid line is the RTA result and the red solid line is the QM result. The classical results are straight lines indicating that magnetic field does not affect the RTA results of longitudinal component of  conductivity. The reason for this behaviour is attributed to the fact that Lorentz force does not do any work in the direction of magnetic field. But contrary to the classical results the QM results show a variation  with magnetic field because of Landau quantization effect. 
 In the lower panel we have shown the plots of $\Delta\sigma / \sigma_\text{RTA}$ for $D^+$ mesons and charm quarks in Fig.~\ref{ec:pll-B}(c) and (d) which basically shows the relative change of QM results with respect to classical results. Here $\Delta\sigma = (\sigma_\text{QM} -\sigma_\text{RTA})\times 100$. From Fig.~\ref{ec:pll-B}(c) and (d) we see that the quantum enhancement of c quark and quantum suppression of $D^+$ meson in conductivity increases with magnetic field. 
By fusing the results of Fig.~\ref{ec:pll-T}(c), (d) and Fig.~\ref{ec:pll-B}(c), (d), we can say that quantum effects become dominant in low $T$ and high $eB$ zone, which is a very well known fact~\cite{Dey:2019axu,Satapathy:2021cjp,Dey:2020awu}.
	\begin{figure}[h]
		\includegraphics[scale = 0.35]{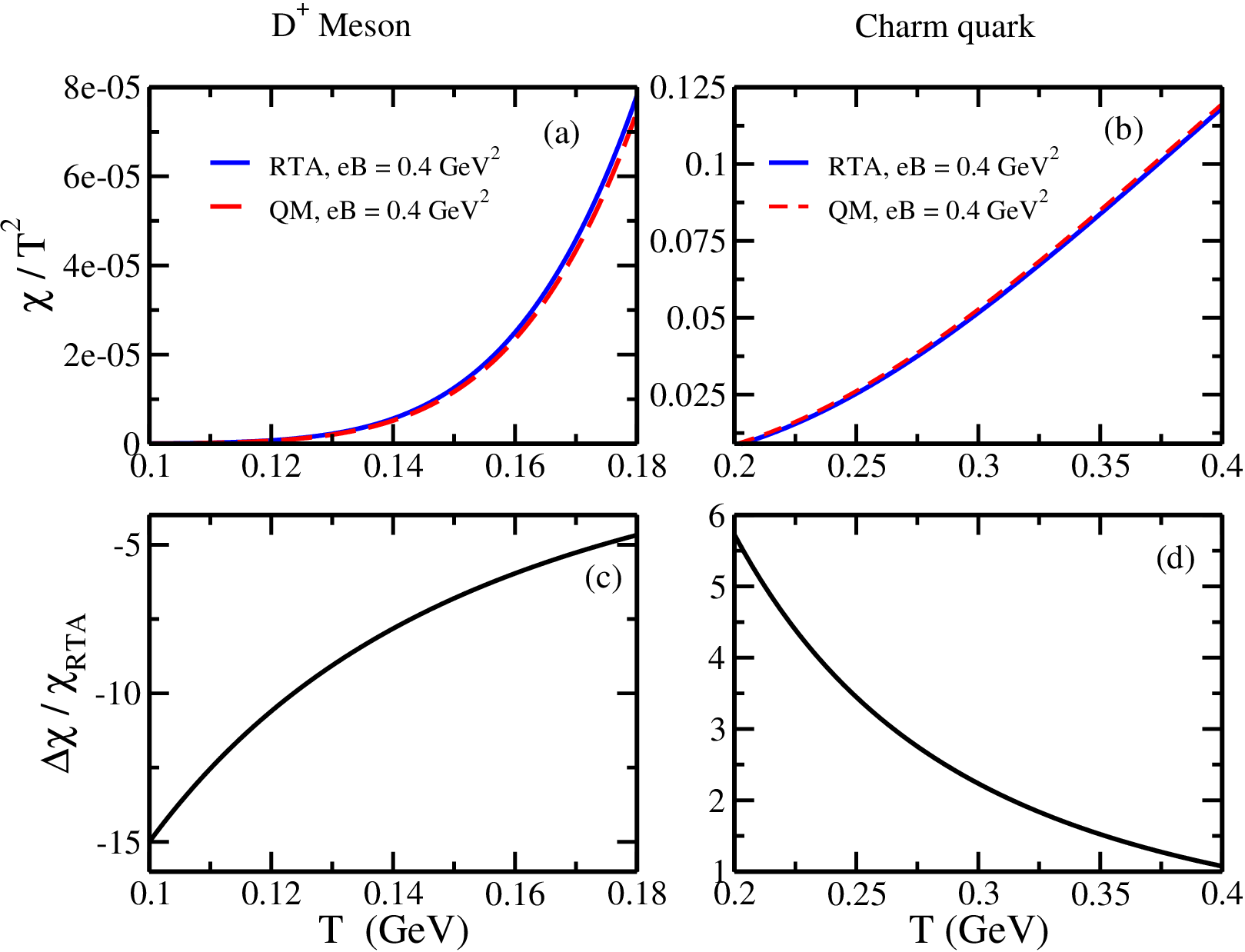}
		\caption{ Plot of susceptibility scaled by $1 / T^2$ shown by (a) $\chi / T^2$ vs $T$ for $D^+$ mesons and (b) $\chi / T^2$ vs $T$ for charm quarks at $eB = 0.4$ GeV$^2$. Relative percentage changes of $\chi / T^2$ of QM results with respect to RTA results for (c) $D^+$ mesons and (d) charm quarks at $eB = 0.4$ GeV$^2$.}
		\label{susc-TB}
	\end{figure}
	\begin{figure}[h]
		\includegraphics[scale = 0.35]{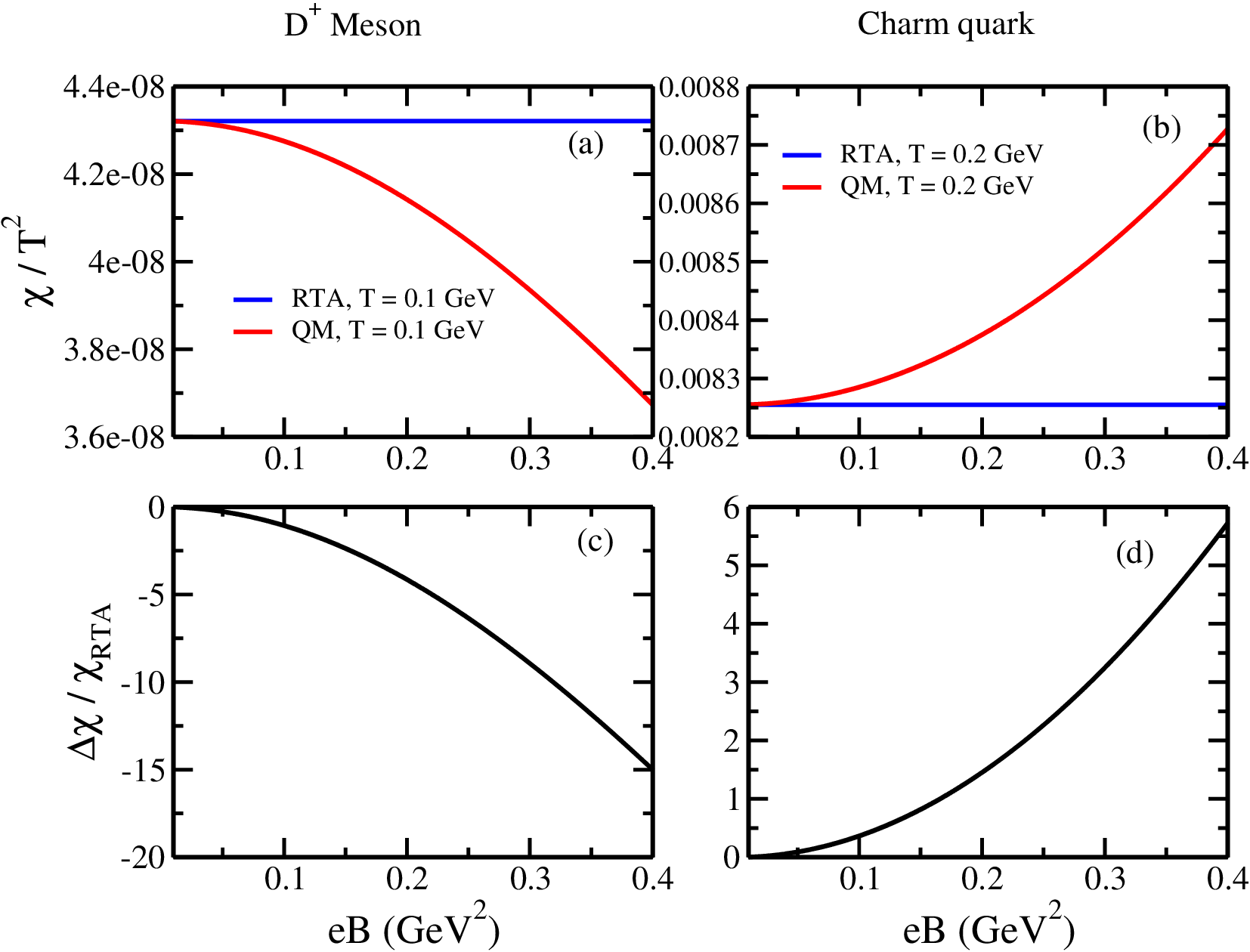}
		\caption{ Plot of susceptibility scaled by $1 / T^2$ shown by (a) $\chi / T^2$ vs $eB$ at $T = 0.1$ GeV for $D^+$ mesons and (b) $\chi / T^2$ vs $eB$ at $T = 0.2$ GeV. Relative percentage changes of $\chi / T^2$ of QM results with respect to RTA results for (c) $D^+$ mesons at $T = 0.1$ GeV and (d) charm quarks at $T = 0.2$ GeV.}
		\label{susc-B}
	\end{figure}
	\begin{figure}[h]
		\includegraphics[scale = 0.35]{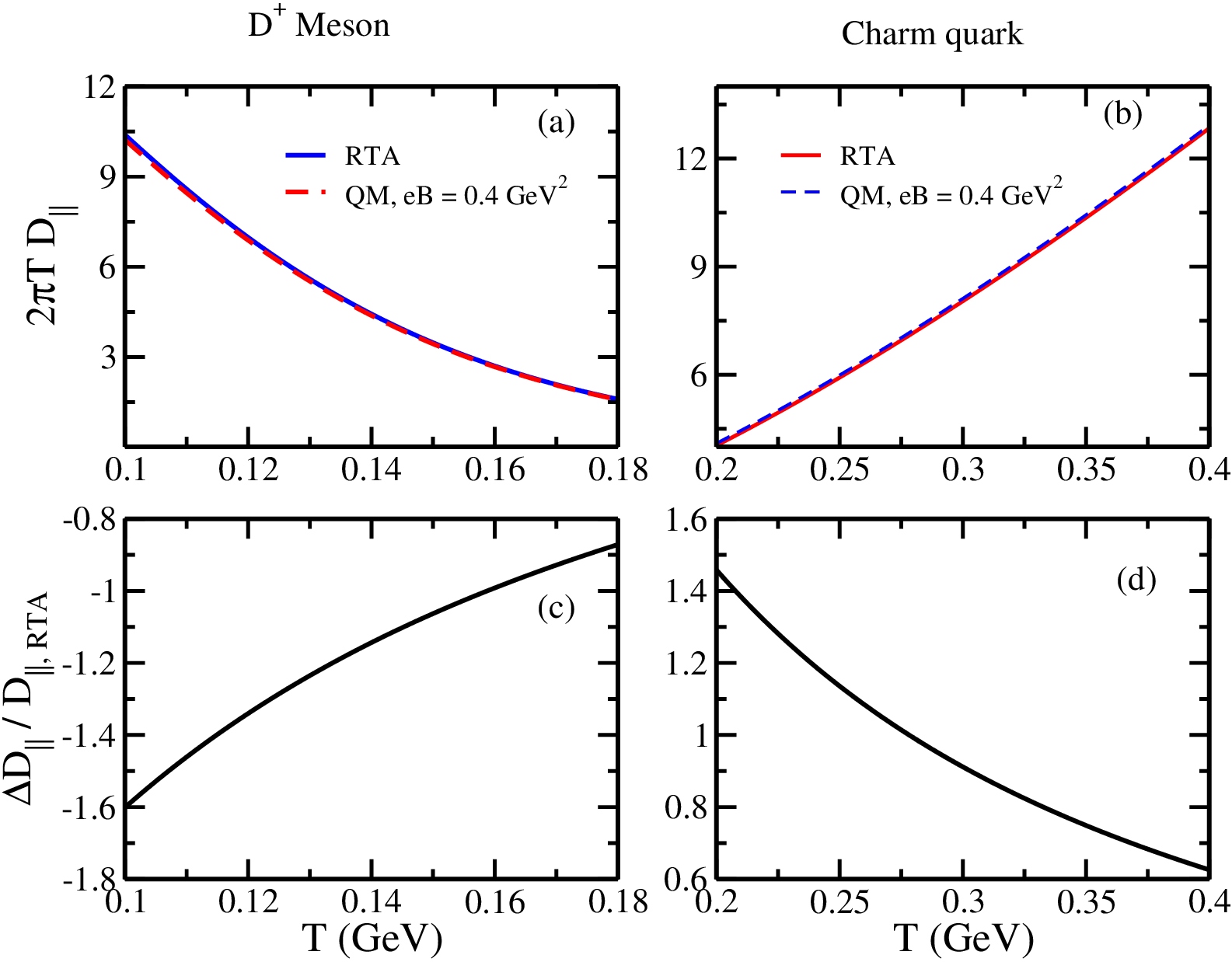}
		\caption{ Normalized values of longitudinal diffusion $D_{\parallel}$ scaled by  $2\pi T $ plotted as a function of $T$ for RTA and QM results for (a) $D^+$ mesons and (b) charm quarks at $eB = 0.4$ GeV$^2$. Relative percentage changes of $D_{\parallel}$ of QM results with respect to RTA results for (c) $D^+$ mesons and (d) charm quarks at $eB = 0.4$ GeV$^2$.}
		\label{D:pll-T}
	\end{figure}
	\begin{figure}[h]
		\includegraphics[scale = 0.35]{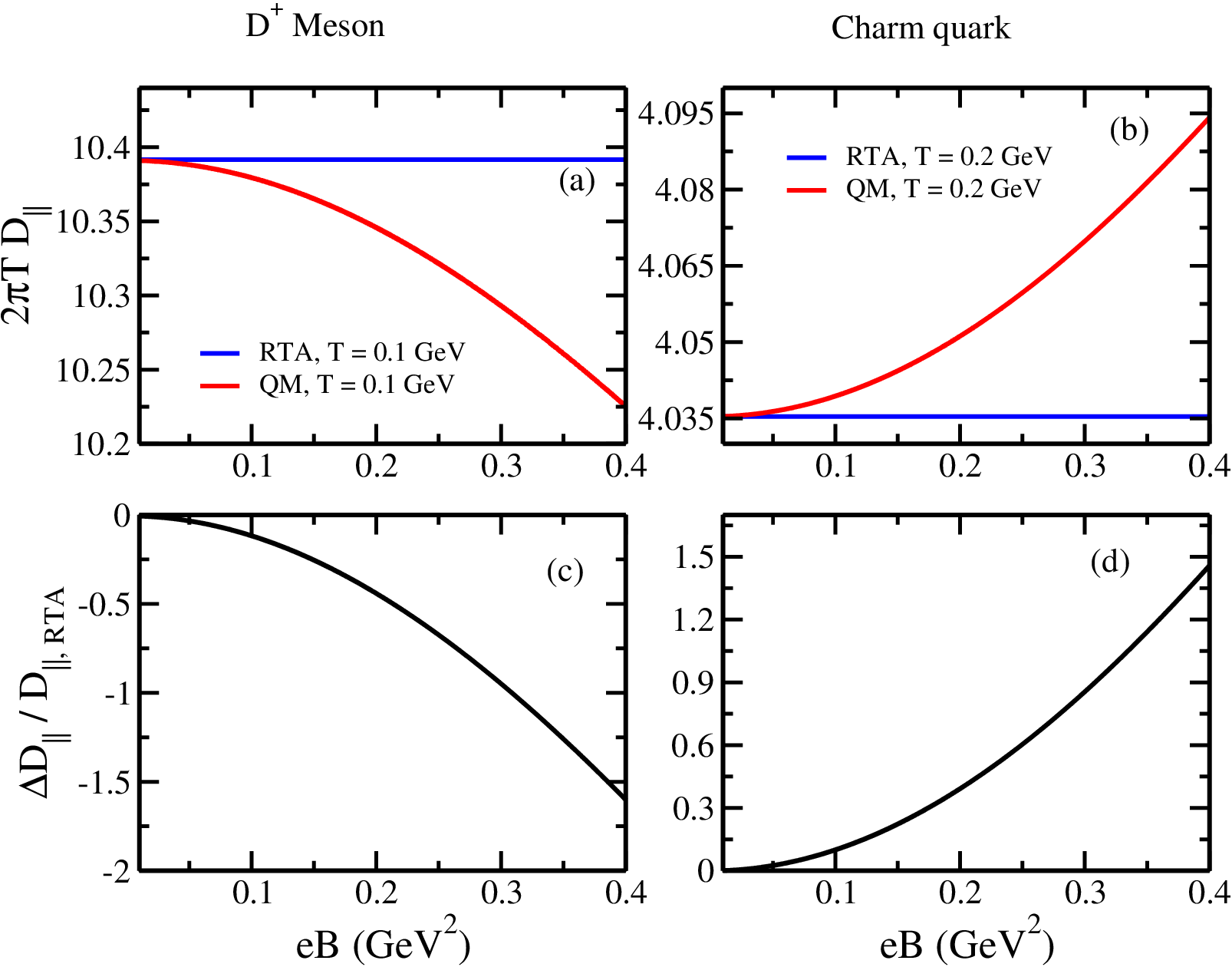}
		\caption{ Normalized values of longitudinal diffusion $D_{\parallel}$ scaled by  $2\pi T $ plotted as a function of $eB$ for RTA and QM results for (a) $D^+$ mesons at $T = 0.1$ GeV and (b) charm quarks at $T = 0.2$ GeV. Relative percentage changes of $D_{\parallel}$ of QM results with respect to RTA results for (c) $D^+$ mesons at $T = 0.1$ GeV and (d) charm quarks at $T = 0.2$ GeV.}
		\label{D:pll-B}
	\end{figure}
	Next, let us go for discussion on heavy quark/meson number susceptibility, given in Eq.~(\ref{f:susc}), which is found to be independent of magnetic field when we don't consider Landau quantization. Through phase space quantization, Eq.~(\ref{f:susc}) can be rewritten as Eq.~(\ref{f:susc:qm}), which we call the QM version of susceptibility, that carries the dependence of magnetic field. In order to study the dependence of susceptibility on $T$ and $B$ we have plotted $\chi / T^2$ vs $T , eB$ in Figs.~\ref{susc-TB} and \ref{susc-B}. 
 In Fig.~\ref{susc-TB}(a) and (b) the blue solid curves are the RTA results and the red dashed curves are the QM results of susceptibility for $D^+$ mesons and charm quarks respectively. In the lower panel i.e., Fig.~\ref{susc-TB}(c) and (d) we have shown the relative changes of $\chi / T^2$ with respect to RTA results where we see that for $D^+$ mesons QM results show a slight suppression whereas for charm quarks there is a slight enhancement as compared to the RTA results. But as the temperature increases the difference decreases, indicating diminished quantum effects at high temperatures, similar to conductivity. We have then shown the variation of susceptibility with magnetic field in Fig.\ref{susc-B}(a) and (b) for $D^+$ mesons and charm quarks at $T = 0.1$ and 0.2 GeV respectively. Here we find that the RTA results of susceptibility, shown by solid blue lines do not vary with magnetic field whereas the QM results shown by solid red lines show a change, deviating away RTA results as the magnetic field increases. The relative changes of $\chi$ for QM results with respect to RTA are plotted in Figs.~\ref{susc-B}(c) and (d) where we see that QM results are suppressed for $D^+$ mesons and enhanced for charm quarks with magnetic field as compared to the RTA results.

	Now we discuss the behaviour of longitudinal component of spatial diffusion $D_{\parallel}$ with $T$ and $B$. 
	The behaviour of spatial diffusion (normalized by $2\pi T$) with temperature has been plotted as $2\pi TD_{\parallel}$ vs $T,eB$ in Figs.~\ref{D:pll-T} and \ref{D:pll-B}. The dependence of spatial diffusion on the temperature for $D^+$ mesons and charm quarks has been shown in Fig.~\ref{D:pll-T}(a) and (b) at $eB = 0.4$ GeV$^2$. 
	RTA results of longitudinal diffusion will be same as diffusion without magnetic field results. So we will get the decreasing diffusion in hadronic temperature domain and increasing diffusion in quark temperature domain with a minimum around transition temperature, as pointed out by Ref.~\cite{Abreu:2011ic}.
	We have also compared the results of diffusion for classical and QM expressions which are given by blue dashed curves and red solid lines in Fig.~\ref{D:pll-T}(a) and (b). Here we see that although the classical and QM curves are very close to each other, their differences are evident as displayed in Fig.~\ref{D:pll-T}(c) and (d) where we plotted $\Delta D / D_\text{RTA}$ with $\Delta D = (D_\text{QM} - D_\text{RTA} )\times 100$. For the $D^+$ mesons there is a small suppression of QM results and for the charm quarks there is a small enhancement of diffusion as compared to RTA results, similar to longitudinal conductivity.
	Next, we have plotted longitudinal diffusion with magnetic field in Fig.~\ref{D:pll-B}. The behaviour of longitudinal diffusion in magnetic field is similar to the behaviour of longitudinal  conductivity. Collecting $\Delta D / D_\text{RTA}$ from Fig.~\ref{D:pll-T}(c), (d) and Fig.~\ref{D:pll-B}(c), (d), we can again find the dominance of quantum aspects in low $T$ and high $eB$ zone. 
	%
	\subsection{Transverse components}
	\begin{figure}[h]
		\includegraphics[scale = 0.35]{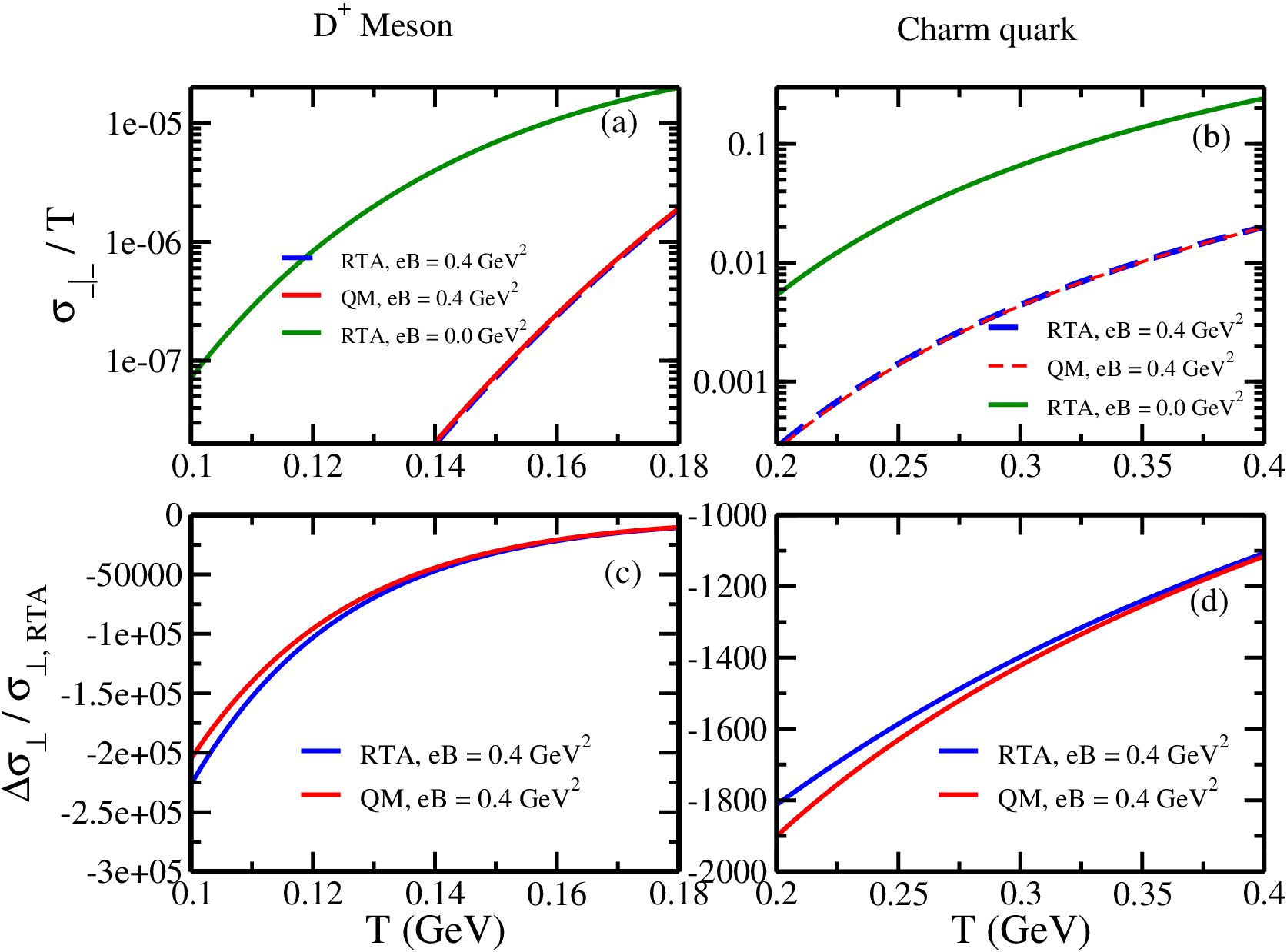}
		\caption{Normalized values of transverse heavy quark conductivity $\sigma_{\perp} / T$ plotted as a function of $T$ for RTA and QM results for (a) $D^+$ mesons and (b) charm quarks at $eB = 0.4$ GeV$^2$. Relative percentage changes of $\sigma_{\perp}$ of QM results with respect to RTA results at $B = 0$ for (a) $D^+$ mesons and (b) charm quarks at $eB = 0.4$ GeV$^2$.  }
		\label{ec:perp-T}
	\end{figure}
	\begin{figure}[h]
		\includegraphics[scale = 0.35]{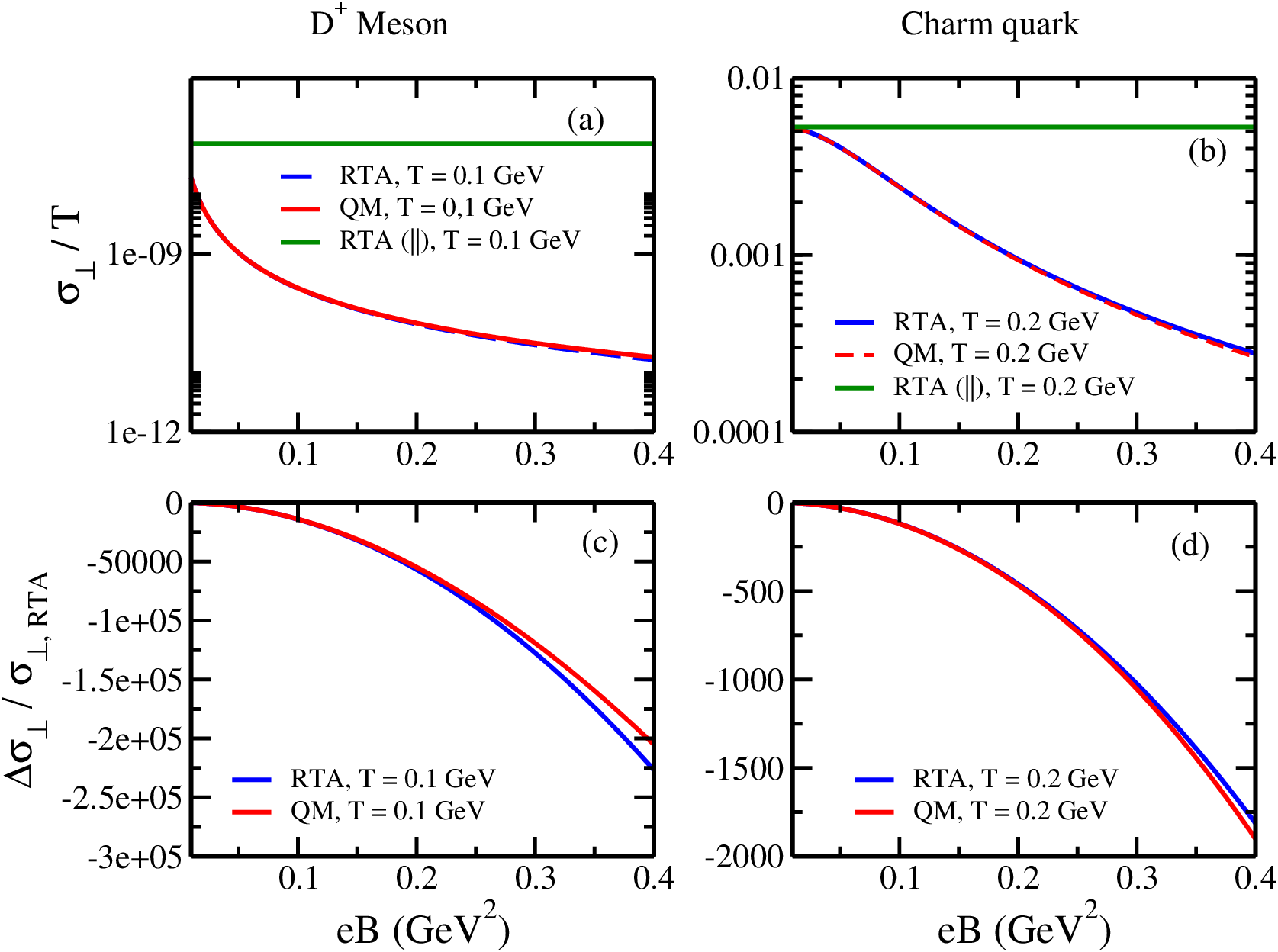}
		\caption{ Normalized values of transverse heavy quark conductivity $\sigma_{\perp} / T$ plotted as a function of $eB$ for RTA and QM results for (a) $D^+$ mesons at $T = 0.1$ GeV and (b) charm quarks at $T = 0.2$ GeV. Relative percentage changes of $\sigma_{\perp}$ of QM results with respect to RTA results at $B = 0$ for (c) $D^+$ mesons at $T = 0.1$ GeV (d) charm quarks at $T = 0.2$ GeV. }
		\label{ec:perp-B}
	\end{figure}
	Let us now discuss the results of transverse components of transport coefficients with temperature and magnetic field. As mentioned before in order to understand the behaviour of diffusion we have to study  conductivity first as a reference.  Transverse  conductivity ($\sigma_{\perp}$) has been studied as a function of temperature and magnetic field in Figs.~(\ref{ec:perp-T}) and (\ref{ec:perp-B}) by plotting the dimensionless quantity $\sigma_{\perp}/T$. The behaviour of $\sigma_{\perp}$ with temperature for $D^+$ mesons and charm quarks has been shown in Fig.~\ref{ec:perp-T}(a) and (b) at $eB = 0.4$ GeV$^2$. 
	The blue dashed curve are the classical results and the red curves (solid line for $D^+$ mesons and dashed lines for charm quarks) are the QM results. 
	In order to understand the relative behaviour of classical and QM results we have plotted $\Delta\sigma / \sigma_\text{RTA}$ in Fig.~\ref{ec:perp-T}(c) and (d), where $\Delta\sigma_\perp = (\sigma^i_\perp - \sigma_\text{B = 0})\times 100$ where $\text{i} = {\text{RTA, QM}}$. Negative values of $\Delta\sigma_\perp$ indicates that perpendicular component of conductivity will be reduced at finite magnetic field from its values at $B=0$. This fact is also connected with standard anisotrpic nature of transport
    coefficients in prersence of magnetic field, where its perpendicular component become smaller than its parallel 
    component~\cite{Dey:2019axu,Satapathy:2021cjp,Dey:2020awu}. This is because the transverse/perpendicular relaxation time $\tau_c^\perp = \tau_c / \big(1 + \frac{\tau_c^2e^2B^2}{\omega^2}\big)$ become smaller than longitudinal/parallel relaxation time $\tau_c^\parallel=\tau_c$.
We notice from Fig.~\ref{ec:perp-T}(c) that the RTA results are more suppressed as compared to QM results with respect to $B = 0$ results for $D^+$ mesons and an opposite trend is observed for charm quarks in Fig.~\ref{ec:perp-T}(d). 
	The reason can be analyse mathematically as follows. Longitudinal and transverse components have two different integrands, which are integrated over the (Landau) quantized phase space. As an outcomes of the integration, we are getting two different directional results with respect to without Landau quantization or RTA results. One is larger and another become smaller with respect to RTA results and accordingly quantum enhancement in $D^+$ meson and suppression in c quark are found. 
	\begin{figure}[h]
		\includegraphics[scale = 0.35]{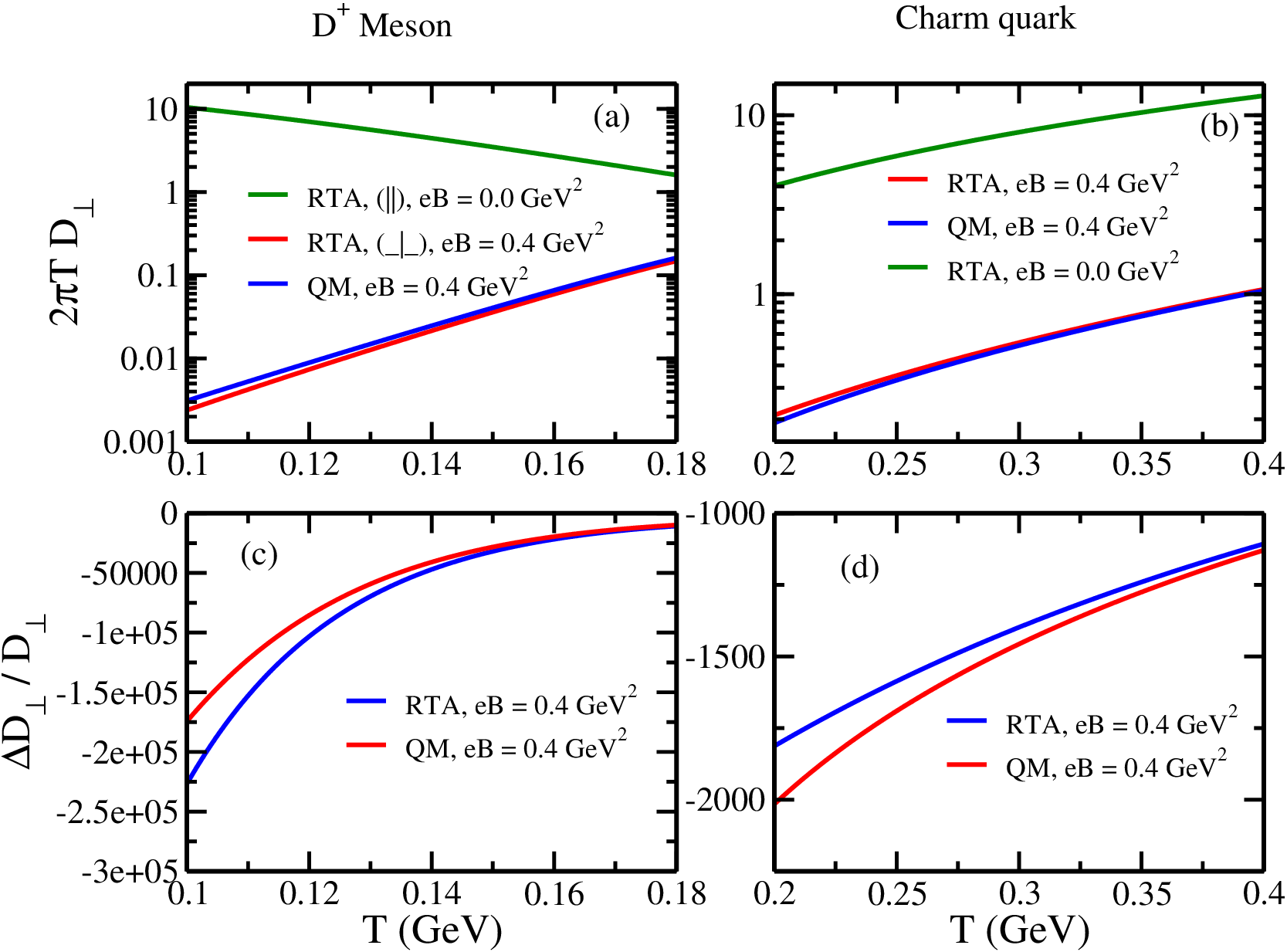}
		\caption{Normalized values of transverse diffusion $D_{\perp}$ scaled by  $2\pi T $ plotted as a function of $T$ for RTA and QM results for (a) $D^+$ mesons and (b) charm quarks at $eB = 0.4$ GeV$^2$. Relative percentage changes of $D_{\perp}$ of QM results with respect to RTA results at $B = 0$ for (c) $D^+$ mesons and (d) charm quarks at $eB = 0.4$ GeV$^2$.}
		\label{D:perp-T}
	\end{figure}
	\begin{figure}[h]
		\includegraphics[scale = 0.35]{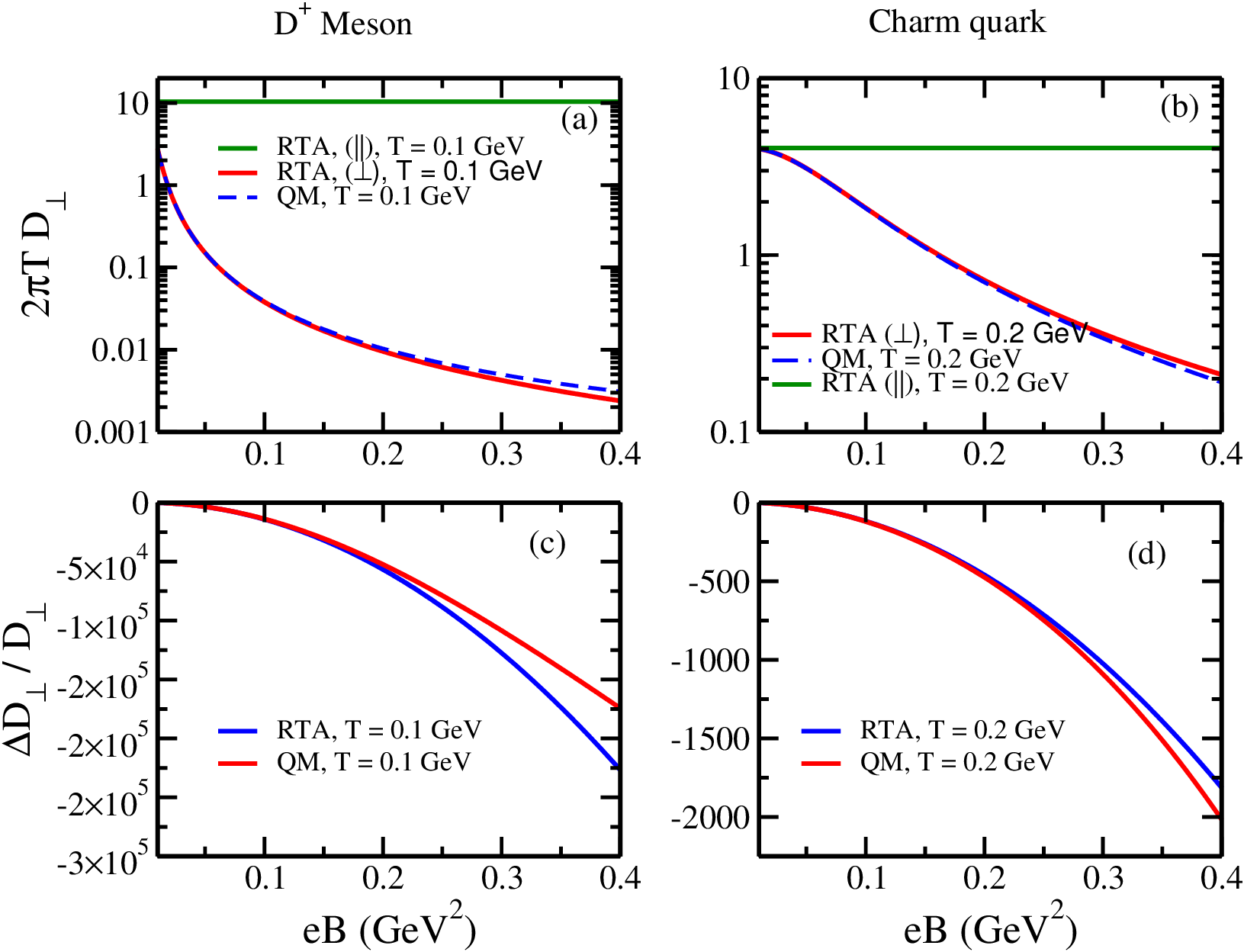}
		\caption{ Normalized values of transverse diffusion $D_{\perp}$ scaled by  $2\pi T $ plotted as a function of $eB$ for RTA and QM results for (a) $D^+$ mesons at $T = 0.1$ GeV and (b) charm quarks at $T = 0.2$ GeV. Relative percentage changes of $D_{\parallel}$ of QM results with respect to RTA results at $B = 0$ for (c) $D^+$ mesons at $T = 0.1$ GeV and (d) charm quarks at $T = 0.2$ GeV.}
		\label{D:perp-B}
	\end{figure}

	Next, in Fig.~\ref{ec:perp-B} (a) and (b), we have plotted $\sigma_{\perp}/ T$ against $eB$-axis, which shows a decreasing pattern due to the term $1/(1 + \frac{\tau_c^2e^2B^2}{\omega^2})$. Difference between QM and RTA results are also shown through the quantity $\Delta\sigma / \sigma_\text{RTA}$ in
	Fig.~\ref{ec:perp-B} (c) and (d). 
 We observe in Fig.~\ref{ec:perp-B}(c) and (d) that QM results dominate over RTA results for $D^+$ mesons but an opposite effect is observed for charm quarks.
  Fig.~\ref{ec:perp-T}(c), (d) and Fig.~\ref{ec:perp-B}(c), (d) collectively reveal that the difference between $\Delta\sigma_\perp$ of RTA and QM curves decreases with temperature and increase with magnetic field for both $D^+$ mesons and charm quarks. It is again reflecting the dominance of quantum effect in low $T$ and high $eB$ domain.

	Next, let us come to the transverse component of diffusion ($D_{\perp}$) and the behaviour with temperature and magnetic field. 
	To understand the difference with respect to isotropic case results we have included the plots of $B=0$ in Fig.~\ref{D:perp-T}(a) and (b) shown by green solid lines. The RTA and QM results at $eB = 0.4$ GeV$^2$ have been shown by blue and red solid lines. Clearly from Fig.~\ref{D:perp-T}(a) for the $D^+$ mesons we see that the $B = 0$ results show a decreasing behaviour whereas as the $B\neq 0$ results exhibit an opposite behaviour. For the charm quarks in Fig.~\ref{D:perp-T}(b) both the $B= 0$ and $B\neq 0$ results show an increasing behaviour with temperature.
	%
	To differentiate and quantify the relative changes of the transverse spatial diffusion from the $B = 0$ results we have plotted $\Delta D_{\perp} / D_{\perp}$ in Fig.~\ref{D:perp-T}(c) and (d) where $\Delta D = ( D_\text{i} - D_\text{B = 0} )\times 100$ for $\text{i} =\{\text{RTA, QM}\} $. 
	The behaviour of transverse diffusion with magnetic field is shown in Fig.~\ref{D:perp-B}.
	The isotropic or $B = 0$ results are shown by green straight lines, the RTA results and QM results at $T = 0.1$ GeV and $T = 0.2$ GeV for $D^+$ mesons and charm quarks are shown by solid red and blue dashed lines respectively in Fig.~\ref{D:perp-B}(a) and (b). 
	In Fig.\ref{D:perp-B}(c) and (d) we have plotted $\Delta D_\perp / D_\perp$ vs $eB$, where $\Delta D = D_\perp^\text{i} - D_\text{B = 0}\times 100$, $\text{i} = \{\text{RTA, QM}\}$. The blue solid lines (RTA) and red solid lines (QM) merge at low magnetic field but grow apart at high magnetic field. Difference between blue (RTA) and red (QM) curves in Fig.~\ref{D:perp-T}(c), (d) and Fig.~\ref{D:perp-B}(c), (d) increases as $T$ decreases and $eB$ increases, which is again supporting the quantum effect dominance in low $T$ and high $eB$ zone.  
	
	At the end, if we focus only on the spatial diffusion curve as a function of temperature, then we notice that our expected $2\pi TD$ vs $T$ curve is modified in
	presence of magnetic field. In the absence of magnetic field, $2\pi TD$ is expected to decrease in hadronic temperature domain and increase in quark temperature domain.
	This trend remains similar for the longitudinal component of diffusion in presence of magnetic field with slight suppression for $D^+$ mesons and enhancement for charm quarks due to Landau quantization. But for the transverse component this trend changes because of anisotropic factors.

	\section{Summary and Conclusion}
	\label{summary}

We have first build relaxation time approximation (RTA) framework for heavy quark and heavy meson conductivity calculations, where
there relaxation times are tuned from the knowledge of earlier works on their spatial diffusion estimations. According to Einstein's diffusion relation, heavy quark and heavy meson conductivity is equal to their corresponding spatial diffusion times susceptibility.   
After tuning our estimation in absence of magnetic field, we have extended RTA framework at finite magnetic field, where parallel and perpendicular components of diffusion and conductivity components for heavy quark and meson are introduced in the picture.

In a short summary, if you denote without magnetic field RTA results as RTA($B=0$), finite magnetic field RTA results as RTA($B\neq 0$) and finite magnetic field RTA with inclusion of Landau quantization as QM($B\neq 0$), then our outcomes can be addressed in bullet points as follows :
\begin{itemize}
    \item For susceptibility, longitudinal conductivity, and longitudinal diffusion component of $D^+$ meson, RTA($B=0)=$ RTA($B\neq 0)>$QM$(B\neq 0)$.
    \item For susceptibility, longitudinal conductivity, and longitudinal diffusion component of $c$ quark, QM$(B\neq 0)>$ RTA($B=0)=$ RTA($B\neq 0)$.
    \item For transverse conductivity and diffusion component of $D^+$ meson, RTA($B=0)>$ QM$(B\neq 0)>$ RTA($B\neq 0)$.
    \item For transverse conductivity and diffusion component of $c$ quark, RTA($B=0)>$ RTA($B\neq 0)>$QM$(B\neq 0)$.
    \item Effect of Landau quantization in all quantities - susceptibility, conductivity and diffusion are prominent at low temperature and high magnetic field domain.
\end{itemize}

Our results may be projected in future for some phenomenological results of heavy meson suppression, where we can expect those anisotropic effect (phenomenological results along parallel and perpendicular may be different) and quantum effect. Although, for the realistic case of time-varying magnetic field, the actual effect of these quantum-anisotropy can be unfolded after doing the future research.
 
	
	\section*{Acknowledgements}
	The authors thank Chitrasen Jena for beginning level collaborative discussions.
	Sudipan De and Sarthak Satapathy acknowledge financial support from the DST INSPIRE Faculty research grant (IFA18-PH220), India.

	
	\appendix
	
	\section{ Heavy quark or heavy meson conductivity from Relativistic Boltzmann equation}
	\label{sec:ECB}
	In this section we derive the heavy quark or heavy meson conductivities in the presence of magnetic field from Relativistic Boltzmann equation(RBE). 
	Let us consider a background magnetic field $\vec{B}$
	pointing in the $z$-direction and heavy quark (c)
	or meson ($D^+$) with charm chemical potential $\mu_c$ and mass $m$, which can have a conduction and diffusion due to
	gradient of $\mu_c$. One can connect macroscopic and microscopic definition charm quark current density owing to which
	the dissipative current density $J^i$ can be expressed as 
	\bea
	J^i &=& \sigma^{ij}\nabla_j\mu_c = g\int\frac{d^3k}{(2\pi)^3}\frac{p^i}{\omega}~\delta f
	\label{ohm} 
	\eea
	where $\sigma^{ij}$ is the heavy quark conductivity tensor, $\nabla_j$ is the spatial derivative, 
	$\delta f$ is the deviation of the distribution function $f$ from equilibrium and
	$g$ is degeneracy factor of heavy quark or heavy meson. Considering 2 for spin degeneracy, 3 for color degeneracy,
	we get $g=2\times 3=6$ for $c$ quark while for $D^+$ meson, we will get $g=1$ from spin degeneracy only.
	In order to study the transport properties of this system we make use of the Relativistic Boltzmann equation(RBE), which is given by 
	\bea
	\frac{\partial f}{\partial t} + \frac{\vec{p}}{\omega}\frac{\partial f}{\partial \vec{x}} + \frac{\partial \vec{p}}{\partial t}\frac{\partial f}{\partial \vec{p}} = I[\delta f] = -\frac{\delta f}{\tau_c}~~.
	\label{RBE-1}
	\eea 
	
	The first term of Eq.\ref{RBE-1} does not contribute to the calculation of heavy quark conductivity as we have not considered the time dependency in $\delta f$. The second and third term survive. The second term of the RBE is evaluated as follows. 
	Assuming $\nabla_i T = 0$ and keeping $\nabla_i\mu_c \neq 0$ we get
	\be
	\frac{\vec{p}}{\omega}\frac{\partial f}{\partial \vec{x}} 
	= \frac{1}{\omega}\frac{\partial f_0}{\partial \omega}\vec{p}.\vec{\nabla}\mu_c~~.
	\label{RBE-11}
	\ee  
	
	The third term is the force term appeared due to Lorentz force in the presence of an external magnetic field $\vec{B}$ in the $\hat{z}$ direction, which is evaluated as
	\bea
	\frac{\partial \vec{p}}{\partial t}\frac{\partial f}{\partial \vec{p}}&=&q\Big(  \frac{\vec{p}}{\omega}\times \vec{B}\Big) \cdot \frac{\partial (f_0 + \delta f)}{\partial \vec{p}}  \nn \\
	&=& q \Big(  \frac{\vec{p}}{\omega}\times \vec{B}\Big)\cdot \frac{\vec p}{\omega}\frac{\partial f_0}{\partial \omega} + q\Big(  \frac{\vec{p}}{\omega}\times \vec{B}\Big) \cdot \frac{\partial (\delta f)}{\partial \vec p}  \nn \\
	&=& 0 + \frac{1}{\tau_B}(\vec{p}\times \hat{b})\cdot \frac{\partial (\delta f)}{\partial \vec p}~,
	\nn \\
	\label{3rd-term}
	\eea
	where $\tau_B = \frac{\om}{q|\vec B|}$ 
	and $\hat{b} = \frac{\vec{B}}{|\vec{B}|}$ is the unit vector along the direction of magnetic field. 
	The deviated part of the distribution function is driven by magnetic field and the gradient of charge chemical potential.
	An ansatz for the change in the distribution function can be considered as \[ \delta f = \vec{p} \cdot \vec{F}\frac{\partial f_0}{\partial \omega}~, \] where $\vec{F}$ is a force due to mangetic field and gradient of chemical potential.
	Now, Eq.~\ref{3rd-term} becomes
	\bea
	\frac{\partial \vec{p}}{\partial t}\frac{\partial f}{\partial \vec{p}}
	&=& \frac{1}{\tau_B}(\vec{p}\times \hat{b})\cdot \vec{F}\frac{\partial f_0}{\partial \omega}
	\label{RBE-13}
	\eea
	%
	We can rewrite the RBE with the help of Eqs.~(\ref{RBE-11}) and (\ref{RBE-13}) as 
	\bea
	\frac{1}{\omega}\frac{\partial f_0}{\partial \omega}\Big[ \vec{p}.\vec{\nabla}\mu_c + \vec{p}.\big(\hat b \times \vec{F}\big)  \Big] = -\frac{\vec{p}.\vec{F}}{\tau_c}\frac{\partial f_0}{\partial \omega}~~.
	\label{RBE-2}
	\eea 
	A general expression of  $\vec{F}$ 
	is given by $$\vec{F} = \alpha \hat{\mu}_c + \beta\hat{b} + \gamma\big(\hat{\mu}_c\times\hat{b}\big)$$ 
	where $\hat{\mu}_c = \frac{\vec{\nabla}\mu_c}{|\vec{\nabla}\mu_c|}$ is the unit vector along the gradient of the chemical potential.
	By substituting the expression of $\vec{F}$ in Eq.\ref{RBE-2}, cancelling the common terms and retaining the coefficients of $\hat{\mu}_c, \hat{b}$ and $\hat{\mu}_c\times\hat{b}$ we get 
	\bea
	\frac{|\vec{\nabla}\mu_c|}{\omega} + \frac{1}{\tau_B}\Big[ -\alpha(\hat{\mu}_c\times \hat{b}) + \gamma\Big\{ \hat{\mu}_c - \hat{b}\big(\hat{b}.\hat{\mu}_c\big) \Big\}\Big] = -\frac{1}{\tau_c}\Big[ \alpha\hat{\mu}_c + \beta \hat{b} + \gamma (\hat{b}\times \hat{\mu}_c)\Big]~~.
	\label{RBE-3}
	\eea 
	
	From Eq.\ref{RBE-3} the coefficients $\alpha, \beta,\gamma$ are given by
	\bea
	\alpha &=& -\frac{\tau_c\big|\vec{\nabla}\mu_c\big|}{\omega} \Bigg[\frac{1}{1 + \frac{\tau_c^2}{\tau_B^2}}\Bigg] \\
	\beta &=& -\frac{\tau_c\frac{\tau_c^2}{\tau_B^2}\big|\vec{\nabla}\mu_c\big|(\hat{b}.\hat{\mu}_c)}{\omega} \Bigg[\frac{1}{1 + \frac{\tau_c^2}{\tau_B^2}}\Bigg] \\
	\gamma &=& -\frac{\tau_c^2\big|\vec{\nabla}\mu_c\big|}{\tau_B~~\omega}\Bigg[\frac{1}{1 + \frac{\tau_c^2}{\tau_B^2}}\Bigg] ~~~.
	\eea 
	
	where $\tau_B = \omega / eB$ is the inverse of cyclotron frequency.
	By substituing $\alpha, \beta, \gamma$ in $\delta f$ we get
	\bea
	\delta f = -\frac{\tau_c}{\omega \Big(1 + \frac{\tau_c^2}{\tau_B^2}\Big)}\Bigg[\delta_{ij} - \frac{\tau_c}{\tau_B}\epsilon_{ijk}b_k + \frac{\tau_c^2}{\tau_B^2}b_ib_j\Bigg]\big|\vec{\nabla}\mu_c\big|_ip_j\frac{\partial f_0}{\partial \omega}~~.
	\eea 
	Substituting the expression of $\delta f$ in the macroscopic expression of the dissipative current density we get
	
	\bea
	J^i &=& \sigma^{ij}\nabla_j\mu_c = g \int \frac{d^3p}{(2\pi)^3}\frac{p^i}{\omega}\delta f\\
	\sigma^{ij} &=& \frac{1}{T}\int\frac{d^3p}{(2\pi)^3}\Bigg(\frac{\vec{p}}{\omega}\Bigg)^2\frac{\tau_c}{1 + \frac{\tau_c^2}{\tau_B^2}}\Bigg[ \delta^{ij} - \frac{\tau_c}{\tau_B}\epsilon^{ijk}b_k  + \frac{\tau_c^2}{\tau_B^2}b^ib^j\Bigg]f_0(1\mp f_0)~~.
	\eea  
	In the above equation, $\mp$ stand for $c$ quark with $g=6$ and $D^+$ meson with $g=1$ respectively. 
	

	
	Till now, the conductivity tensor $\sigma^{ij}$ has been derived via RTA formalism where the classical definition of cyclotron frequency is still present and there is no effect of Landau quantization of energies. Taking these points into account we call our obtained results as classical results. We can now calculate the $\parallel,\perp$ and $\times$ component of conductivity~\cite{Dash:2020vxk} as 
	\bea
	\sigma_\parallel^{\text{CM}} &=& g\beta\int\frac{d^3k}{(2\pi)^3}\frac{(k_z)^2}{\omega^2}\tau_c^\parallel f_0 \big[1\mp f_0\big]
	\nn\\
	&=& \frac{g\beta}{3}\int\frac{d^3k}{(2\pi)^3}\frac{(k)^2}{\omega^2}\tau_c^\parallel f_0 \big[1\mp f_0\big]
	\label{ec-6}\\
	\sigma_\perp^{\text{CM}} &=&  g\beta\int\frac{d^3k}{(2\pi)^3}\frac{(k_{x})^2}{\omega^2}\tau_c^\perp f_0 \big[1\mp f_0\big]
	\nn\\
	&=&  \frac{g\beta}{3}\int\frac{d^3k}{(2\pi)^3}\frac{(k)^2}{\omega^2}\tau_c^\perp f_0 \big[1\mp f_0\big]
	\label{ec-7}\\
	\sigma_\times^{\text{CM}}&=& g\beta\int\frac{d^3k}{(2\pi)^3}\frac{(k_xk_y)}{\omega^2} \tau_c^\times f_0\big[1 \mp f_0\big]
	\nn\\ 
	&=& \frac{g\beta}{3}\int\frac{d^3k}{(2\pi)^3}\frac{(k)^2}{\omega^2} \tau_c^\times f_0\big[1 \mp f_0\big]
	\label{ec-8}
	\eea 
	where the superscript CM denotes the classical or RTA results and $\tau^\parallel = \tau_c$, $\tau^\perp = \frac{\tau_c}{1 + \frac{\tau_c^2}{\tau_B^2}}$, $\tau_c^\times = \frac{\tau_c^2/\tau_B}{1 + \frac{\tau_c^2}{\tau_B^2}}$, $\tau_c$ is the relaxation time, $\tau_B = \frac{\omega}{qB}$ is the inverse of the cyclotron frequency. For $c$ quark, $g=6$ and Fermi-Dirac distribution have to be considered, whereas, for $D^+$ meson, $g=1$ and Bose-Einstein distribution have to be taken.
	
	One can easily check that in the limit of $eB\rightarrow 0$, $\sigma^{\parallel}$ and $\sigma^{\perp}$ become
	same and $\sigma^{\parallel}$ is basically same as the expression of $c$ quark and $D^+$ meson conductivity in absence of magnetic field
	\be 
	\sigma=\frac{g\beta}{3}\int\frac{d^3k}{(2\pi)^3}\frac{(k_z)^2}{\omega^2}\tau_c f_0\big[1\mp f_0\big]~.
	\ee 
	
	Since energy gets quantized in the presence of magnetic field, the RTA results for $c$ quark can be modified to include a summation over Landau levels by using the approximation given by 
	$$2\int\frac{d^3k}{(2\pi)^3} = 2\int\int\frac{d^2k_{\perp}}{(2\pi)^2}\frac{dk_z}{2\pi} \to \sum_{l=0}^\infty\frac{g_l qB}{2\pi}\int_{-\infty}^{+\infty}\frac{dk_z}{2\pi}~~,$$
	where spin degeneracy factor 2 of CM expression will be replaced by $g_l = 2-\delta_{l,0}$ in QM expression and we have used the approximation $$k_x^2 \approx k_xk_y \approx k_y^2 \approx \frac{k_x^2+k_y^2}{2} = lqB.$$
	The expressions of $c$ quark conductivity in the quantum theoretical expression result are given by 
	\bea
	\sigma_\perp^\text{QM} &=& \frac{3}{T}\sum_{l=0}^\infty g_l\frac{qB}{2\pi}\int_{-\infty}^{+\infty}\frac{dk_z}{2\pi}\frac{lB}{\omega_l^2}\tau^\perp f_0(1-f_0)
	\label{ecqm:perp-2}\\
	\sigma_\parallel^\text{QM} &=& \frac{3}{T}\sum_{l=0}^\infty g_l\frac{qB}{2\pi}\int_{-\infty}^{+\infty}\frac{dk_z}{2\pi}\frac{k_z^2}{\omega_l^2} \tau^\parallel f_0(1-f_0) \\
	\label{ecqm:pll-2}
	\sigma_\times^\text{QM} &=&\frac{3}{T}\sum_{l=0}^\infty g_l\frac{qB}{2\pi}\int_{-\infty}^{+\infty}\frac{dk_z}{2\pi}\frac{lB}{\omega_l^2}\tau^\times f_0(1-f_0)
	\label{ecqm:hall-2}
	\eea

	where the superscript QM denotes quantum theoretical results and $\omega_l = \sqrt{k_z^2 + m^2 + 2lqB}$ is the Landau quantized energy with $q=\frac{2}{3}e$. 
	
	Similarly $D^+$ meson follows the bosonic Landau quantization rules
	$$\int\frac{d^3k}{(2\pi)^3} \to \sum_{l=0}^\infty\frac{qB}{2\pi}\int_{-\infty}^{+\infty}\frac{dk_z}{2\pi}~~,$$
	and
	$$k_x^2 \approx k_xk_y \approx k_y^2 \approx \frac{k_x^2+k_y^2}{2} = \Big(l+1/2\Big)qB.$$
	The expressions of $D^+$ meson conductivity in the quantum theoretical expression result are given by 
	\bea
	\sigma_\perp^\text{QM} &=& \frac{1}{T}\sum_{l=0}^\infty \frac{qB}{2\pi}\int_{-\infty}^{+\infty}\frac{dk_z}{2\pi}\frac{(l+1/2)qB}{\omega_l^2}\tau^\perp f_0(1+f_0)
	\label{ecqm:perp-2}\\
	\sigma_\parallel^\text{QM} &=& \frac{1}{T}\sum_{l=0}^\infty \frac{qB}{2\pi}\int_{-\infty}^{+\infty}\frac{dk_z}{2\pi}\frac{k_z^2}{\omega_l^2} \tau^\parallel f_0(1+f_0) \\
	\label{ecqm:pll-2}
	\sigma_\times^\text{QM} &=&\frac{1}{T}\sum_{l=0}^\infty \frac{qB}{2\pi}\int_{-\infty}^{+\infty}\frac{dk_z}{2\pi}\frac{(l+1/2)qB}{\omega_l^2}\tau^\times f_0(1+f_0)
	\label{ecqm:hall-2}
	\eea 
	where the superscript QM denotes quantum theoretical results and $\omega_l = \sqrt{k_z^2 + m^2 + (2l+1)qB}$ is the Landau quantized energy with $q=e$. Reader can find the transformation from Pauli suppression factor $\frac{\partial f_0}{\partial \omega}=-\beta f_0(1-f_0)$ to Bose enhancement factor $\frac{\partial f_0}{\partial \omega}=-\beta f_0(1+f_0)$ for $D^+$ meson.


	\section{Quark number susceptibility}
	\label{QNS}
	
	Creation of heavy quarks in the medium at $t\to 0$, induces a change in chemical potential given by 
	$$\mu (\vec{x}) = \mu_0 + \delta\mu(\vec{x}),$$
	where $\mu_0$ is the chemical potential at $t = 0$ and $\delta\mu(\vec{x})$ is the change in chemical potential. The thermal distribution at $t \to 0$ is given by 
	\bea
	\frac{1}{e^{\beta(E - \mu(\vec{x}))} \mp 1}  = f_0 + f_0\big(1 \pm f_0\big)\frac{\delta\mu(\vec{x})}{T}~~,
	\label{del-f}
	\eea
	where $\pm$ stands for bosons/fermions, $f_0 = \big[e^{\beta(E - \mu_0(\vec{x}))} \mp 1\big]^{-1} $ is the initial thermal distribution function. From Eq.~(\ref{del-f}) the change in distribution function is given by 
	\bea
	\delta f = f_0\big(1 \pm f_0\big)\frac{\delta\mu(\vec{x})}{T}
	\label{delta-f}
	\eea 
	For very short time scales the collisionless Boltzmann equation is given by 
	\bea
	\Big[\frac{\partial}{\partial t} + v^i\frac{\partial}{\partial x^i}\Big]f(\vec{x},\vec{p},t) = 0
	\eea 
	whose solution is given by 
	$$f(\vec{x},\vec{p},t) = f(\vec{x}-\vec{v}t,\vec{p})~~.$$ The fluctuations in number density is given by 
	\bea
	\delta N (\vec{x},t) = \int\frac{d^3p}{(2\pi)^3}\delta f(\vec{x},\vec{p},t)~~
	\label{del-N}
	\eea 
	whose spatial Fourier transform is given by  
	\bea
	\delta N(\vec{k},t) = \frac{1}{T}\int\frac{d^3p}{(2\pi)^3}~e^{-i\vec{k}.\vec{v}t}f_0(1\pm f_0)~\delta \mu(\vec{k}).\nn 
	\eea
	where we have substituted the expression of $\delta f$ given by Eq.~(\ref{delta-f}).
	For small time scales, we expand the exponential and get 
	\bea
	\delta N (\vec{k},t) = \Big[\chi_s(\vec{k}) -\frac{1}{2}t^2k^2\chi_s(\vec{k})\Big\langle\frac{v^2}{3}\Big\rangle\Big]\delta\mu(\vec{k})\nn
	\eea
	where 
	\bea
	\chi_s(\vec{k}) = \frac{\partial N}{\partial\mu_0} = \frac{1}{T}\int\frac{d^3p}{(2\pi)^3}~f_0(1\pm f_0),{\rm~and}~~~\Big\langle\frac{v^2}{3}\Big\rangle = \frac{1}{T\chi_s(\vec{k})} \int\frac{d^3p}{(2\pi)^3}~f_0(1\pm f_0)\frac{\vec{v}^2}{3}
	\eea 
	where $\pm$ stands for bosons/fermions.
	For $c$ quark (fermion) and $D^+$ meson, $\chi$ are given by
	\bea
	\chi = g\beta \int\frac{d^3k}{(2\pi)^3}~f_0(1\mp f_0)~~~.
	\label{susc-1}
	\eea 
	On applying Landau quantization of energies and quantizing the phase space part of the momentum integral~\cite{Dudal:2018rki},
	we obtain
	\be
	\chi = 3\beta\sum_{l=0}^{\infty}\frac{g_l qB}{2\pi}\int_{-\infty}^{+\infty}\frac{dk_z}{2\pi}~f_0(1-f_0)
	\ee 
	for $c$ quark and\\
	
	\be
	\chi = \beta\sum_{l=0}^{\infty}\frac{ qB}{2\pi}\int_{-\infty}^{+\infty}\frac{dk_z}{2\pi}~f_0(1+f_0)
	\ee 
	for $D^+$ meson.   


\begin{thebibliography}{100}
		
		\bibitem{Shuryak:2004cy}
		E.~V.~Shuryak,
		\href{https://doi.org/10.1016/j.nuclphysa.2004.10.022}{Nucl. Phys. A \textbf{750}, 64-83 (2005)}
		[arXiv:hep-ph/0405066 [hep-ph]].
		
		


  \bibitem{Prino:2016cni}
F.~Prino and R.~Rapp,
\href{https://doi.org/10.1088/0954-3899/43/9/093002}{J. Phys. G \textbf{43}, no.9, 093002 (2016)}
[arXiv:1603.00529 [nucl-ex]].
		
		


  \bibitem{Rapp:2018qla}
R.~Rapp, P.~B.~Gossiaux, A.~Andronic, R.~Averbeck, S.~Masciocchi, A.~Beraudo, E.~Bratkovskaya, P.~Braun-Munzinger, S.~Cao and A.~Dainese, \textit{et al.}
\href{https://doi.org/10.1016/j.nuclphysa.2018.09.002}{Nucl. Phys. A \textbf{979}, 21-86 (2018)}
[arXiv:1803.03824 [nucl-th]].
		
		\bibitem{Das:2010tj}
		S.~K.~Das, J.~e.~Alam and P.~Mohanty,
		\href{https://doi.org/10.1103/PhysRevC.82.014908}{Phys. Rev. C \textbf{82}, 014908 (2010)}
		[arXiv:1003.5508 [nucl-th]].
		
		\bibitem{Skokov:2009qp}
		V.~Skokov, A.~Y.~Illarionov and V.~Toneev,
		\href{https://doi.org/10.1142/S0217751X09047570}{Int. J. Mod. Phys. A \textbf{24}, 5925-5932 (2009)}
		[arXiv:0907.1396 [nucl-th]].
		
		\bibitem{Bzdak:2011yy}
		A.~Bzdak and V.~Skokov,
		\href{https://doi.org/10.1016/j.physletb.2012.02.065}{Phys. Lett. B \textbf{710}, 171-174 (2012)}
		[arXiv:1111.1949 [hep-ph]].
		
		\bibitem{Tuchin:2013ie}
		K.~Tuchin,
		\href{https://doi.org/10.1155/2013/490495}{Adv. High Energy Phys. \textbf{2013}, 490495 (2013)}
		[arXiv:1301.0099 [hep-ph]].
		
		\bibitem{Banerjee:2021sjm}
		D.~Banerjee, S.~Paul, P.~Das, A.~Modak, A.~Budhraja, S.~Ghosh and S.~K.~Prasad,
		\href{https://doi.org/10.48550/arXiv.2103.14440}{[arXiv:2103.14440 [hep-ph]]}.
		
		\bibitem{Das:2016cwd}
		S.~K.~Das, S.~Plumari, S.~Chatterjee, J.~Alam, F.~Scardina and V.~Greco,
		\href{https://doi.org/10.1016/j.physletb.2017.02.046}{Phys. Lett. B \textbf{768}, 260-264 (2017)}
		[arXiv:1608.02231 [nucl-th]].
		
		\bibitem{Kharzeev:2007jp}
		D.~E.~Kharzeev, L.~D.~McLerran and H.~J.~Warringa,
		\href{https://doi.org/10.1016/j.nuclphysa.2008.02.298}{Nucl. Phys. A \textbf{803}}, 227-253 (2008)
		[arXiv:0711.0950 [hep-ph]].
		
		\bibitem{Fukushima:2015wck}
		K.~Fukushima, K.~Hattori, H.~U.~Yee and Y.~Yin,
		\href{https://doi.org/10.1103/PhysRevD.93.074028}{Phys. Rev. D \textbf{93}, no.7, 074028 (2016)}
		[arXiv:1512.03689 [hep-ph]].
		
		\bibitem{Finazzo:2016mhm}
		S.~I.~Finazzo, R.~Critelli, R.~Rougemont and J.~Noronha,
		\href{https://doi.org/10.1103/PhysRevD.94.054020}{Phys. Rev. D \textbf{94}, no.5, 054020 (2016)}
		[erratum: \href{https://doi.org/10.1103/PhysRevD.96.019903}{Phys. Rev. D \textbf{96}, no.1, 019903 (2017)}]
		[arXiv:1605.06061 [hep-ph]].
		
		\bibitem{Goswami:2022szb}
		K.~Goswami, D.~Sahu and R.~Sahoo,
		\href{https://doi.org/10.1103/PhysRevD.107.014003}{Phys. Rev. D \textbf{107}, no.1, 014003 (2023)}
		[arXiv:2206.13786 [hep-ph]].
		
		\bibitem{Dudal:2014jfa}
		D.~Dudal and T.~G.~Mertens,
		\href{https://doi.org/10.1103/PhysRevD.91.086002}{Phys. Rev. D \textbf{91}, 086002 (2015)}
		[arXiv:1410.3297 [hep-th]].
		
		\bibitem{Dudal:2018rki}
		D.~Dudal and T.~G.~Mertens,
		\href{https://doi.org/10.1103/PhysRevD.97.054035}{Phys. Rev. D \textbf{97}, no.5, 054035 (2018)}
		[arXiv:1802.02805 [hep-th]].
		
		
		\bibitem{Dey:2019axu}
		J.~Dey, S.~Satapathy, P.~Murmu and S.~Ghosh,
		\href{https://doi.org/10.1007/s12043-021-02148-3}{Pramana \textbf{95}, no.3, 125 (2021)}
		[arXiv:1907.11164 [hep-ph]].
		
		
		\bibitem{Satapathy:2021cjp}
		S.~Satapathy, S.~Ghosh and S.~Ghosh,
		\href{https://doi.org/10.1103/PhysRevD.104.056030}{Phys. Rev. D \textbf{104}, no.5, 056030 (2021)}
		[arXiv:2104.03917 [hep-ph]].
		
		\bibitem{Dey:2020awu}
		J.~Dey, S.~Samanta, S.~Ghosh and S.~Satapathy,
		\href{https://arxiv.org/abs/2002.04434}{[arXiv:2002.04434 [nucl-th]]}.
		
		\bibitem{Dash:2020vxk}
		A.~Dash, S.~Samanta, J.~Dey, U.~Gangopadhyaya, S.~Ghosh and V.~Roy,
		\href{https://doi.org/10.1103/PhysRevD.102.016016}{Phys. Rev. D \textbf{102}, no.1, 016016 (2020)}
		[arXiv:2002.08781 [nucl-th]].
		

  \bibitem{Nam:2012sg}
S.~i.~Nam,
\href{https://doi.org/10.1103/PhysRevD.86.033014}{Phys. Rev. D \textbf{86}, 033014 (2012)}
[arXiv:1207.3172 [hep-ph]].
\bibitem{Hattori:2016cnt}
K.~Hattori and D.~Satow,
\href{https://doi.org/10.1103/PhysRevD.94.114032}{Phys. Rev. D \textbf{94}, no.11, 114032 (2016)}
[arXiv:1610.06818 [hep-ph]].
  
\bibitem{Hattori:2016lqx}
K.~Hattori, S.~Li, D.~Satow and H.~U.~Yee,
\href{https://doi.org/10.1103/PhysRevD.95.076008}{Phys. Rev. D \textbf{95}, no.7, 076008 (2017)}
[arXiv:1610.06839 [hep-ph]].
  
\bibitem{Harutyunyan:2016rxm}
A.~Harutyunyan and A.~Sedrakian,
\href{https://doi.org/10.1103/PhysRevC.94.025805}{Phys. Rev. C \textbf{94}, no.2, 025805 (2016)}
[arXiv:1605.07612 [astro-ph.HE]].
  

  \bibitem{Kerbikov:2014ofa}
B.~O.~Kerbikov and M.~A.~Andreichikov,
\href{https://doi.org/10.1103/PhysRevD.91.074010}{Phys. Rev. D \textbf{91}, no.7, 074010 (2015)}
[arXiv:1410.3413 [hep-ph]].


  
\bibitem{Feng:2017tsh}
B.~Feng,
\href{https://doi.org/10.1103/PhysRevD.96.036009}{Phys. Rev. D \textbf{96}, no.3, 036009 (2017)}.

  
\bibitem{Fukushima:2017lvb}
K.~Fukushima and Y.~Hidaka,
\href{https://doi.org/10.1103/PhysRevLett.120.162301}{Phys. Rev. Lett. \textbf{120}, no.16, 162301 (2018)}
[arXiv:1711.01472 [hep-ph]].
  

  \bibitem{Li:2018ufq}
W.~Li, S.~Lin and J.~Mei,
\href{https://doi.org/10.1103/PhysRevD.98.114014}{Phys. Rev. D \textbf{98}, no.11, 114014 (2018)}
[arXiv:1809.02178 [hep-th]].

  

 \bibitem{Das:2019wjg}
A.~Das, H.~Mishra and R.~K.~Mohapatra,
\href{https://doi.org/10.1103/PhysRevD.99.094031}{Phys. Rev. D \textbf{99}, no.9, 094031 (2019)}
[arXiv:1903.03938 [hep-ph]].

  

\bibitem{Das:2019ppb}
A.~Das, H.~Mishra and R.~K.~Mohapatra,
\href{https://doi.org/10.1103/PhysRevD.101.034027}{Phys. Rev. D \textbf{101}, no.3, 034027 (2020)}
[arXiv:1907.05298 [hep-ph]].
  

  \bibitem{Ghosh:2019ubc}
S.~Ghosh, A.~Bandyopadhyay, R.~L.~S.~Farias, J.~Dey and G.~Krein,
\href{https://doi.org/10.1103/PhysRevD.102.114015}{Phys. Rev. D \textbf{102}, 114015 (2020)}
[arXiv:1911.10005 [hep-ph]].
		%
		%

  \bibitem{Satapathy:2021wex}
S.~Satapathy, S.~Ghosh and S.~Ghosh,
\href{https://doi.org/10.1103/PhysRevD.106.036006}{Phys. Rev. D \textbf{106}, no.3, 036006 (2022)}
[arXiv:2112.08236 [hep-ph]].
		%


  \bibitem{Ghosh:2020wqx}
S.~Ghosh and S.~Ghosh,
\href{https://doi.org/10.1103/PhysRevD.103.096015}{Phys. Rev. D \textbf{103}, 096015 (2021)}
[arXiv:2011.04261 [hep-ph]].
		%

\bibitem{Li:2017tgi}
S.~Li and H.~U.~Yee,
\href{https://doi.org/10.1103/PhysRevD.97.056024}{Phys. Rev. D \textbf{97}, no.5, 056024 (2018)}
[arXiv:1707.00795 [hep-ph]].
  
\bibitem{Nam:2013fpa}
S.~i.~Nam and C.~W.~Kao,
\href{https://doi.org/10.1103/PhysRevD.87.114003}{Phys. Rev. D \textbf{87}, no.11, 114003 (2013)}
[arXiv:1304.0287 [hep-ph]].
  


  \bibitem{Alford:2014doa}
M.~G.~Alford, H.~Nishimura and A.~Sedrakian,
\href{https://doi.org/10.1103/PhysRevC.90.055205}{Phys. Rev. C \textbf{90}, no.5, 055205 (2014)}
[arXiv:1408.4999 [hep-ph]].

  

  \bibitem{Tawfik:2016ihn}
A.~N.~Tawfik, A.~M.~Diab and T.~M.~Hussein,
Int. J. Adv. Res. Phys. Sci. \textbf{3}, 4-14 (2016)
\href{https://arxiv.org/abs/1608.01034}{[arXiv:1608.01034 [hep-ph]]}.


  
\bibitem{Tuchin:2011jw}
K.~Tuchin,
\href{https://doi.org/10.1088/0954-3899/39/2/025010}{J. Phys. G \textbf{39}, 025010 (2012)}
[arXiv:1108.4394 [nucl-th]].
  

\bibitem{Ghosh:2018cxb}
S.~Ghosh, B.~Chatterjee, P.~Mohanty, A.~Mukharjee and H.~Mishra,
\href{https://doi.org/10.1103/PhysRevD.100.034024}{Phys. Rev. D \textbf{100}, no.3, 034024 (2019)}
[arXiv:1804.00812 [hep-ph]].
  
\bibitem{Mohanty:2018eja}
P.~Mohanty, A.~Dash and V.~Roy,
\href{https://doi.org/10.1140/epja/i2019-12705-7}{Eur. Phys. J. A \textbf{55}, no.3, 35 (2019)}
[arXiv:1804.01788 [nucl-th]].
  
		%
		%

\bibitem{Dey:2019vkn}
J.~Dey, S.~Satapathy, A.~Mishra, S.~Paul and S.~Ghosh,
\href{https://doi.org/10.1142/S0218301321500440}{Int. J. Mod. Phys. E \textbf{30}, no.06, 2150044 (2021)}
[arXiv:1908.04335 [hep-ph]].
  
		%
		%
\bibitem{Hattori:2017qih}
K.~Hattori, X.~G.~Huang, D.~H.~Rischke and D.~Satow,
\href{https://doi.org/10.1103/PhysRevD.96.094009}{Phys. Rev. D \textbf{96}, no.9, 094009 (2017)}
[arXiv:1708.00515 [hep-ph]].
  

\bibitem{Huang:2009ue}
X.~G.~Huang, M.~Huang, D.~H.~Rischke and A.~Sedrakian,
\href{https://doi.org/10.1103/PhysRevD.81.045015}{Phys. Rev. D \textbf{81}, 045015 (2010)}
[arXiv:0910.3633 [astro-ph.HE]].
  

\bibitem{Huang:2011dc}
X.~G.~Huang, A.~Sedrakian and D.~H.~Rischke,
\href{https://doi.org/10.1016/j.aop.2011.08.001}{Annals Phys. \textbf{326}, 3075-3094 (2011)}
[arXiv:1108.0602 [astro-ph.HE]].
  


  \bibitem{Agasian:2011st}
N.~O.~Agasian,
\href{https://doi.org/10.1134/S0021364012040029}{JETP Lett. \textbf{95}, 171-175 (2012)}
[arXiv:1109.5849 [hep-ph]].




\bibitem{Agasian:2013wta}
N.~O.~Agasian,
\href{https://doi.org/10.1134/S1063778813100025}{Phys. Atom. Nucl. \textbf{76}, 1382-1386 (2013)}.
  
		%
		%
		\bibitem{Romatschke:2017ejr}
		P.~Romatschke and U.~Romatschke,
		\href{https://doi.org/10.1017/9781108651998}{Cambridge University Press, 2019,
			ISBN 978-1-108-48368-1, 978-1-108-75002-8}
		doi:10.1017/9781108651998
		[arXiv:1712.05815 [nucl-th]].
		
		\bibitem{Teaney}P. Petreczky, D. Teaney,
		Phys. Rev. D 73 (2006) 014508
		
		\bibitem{Laine:2016hma}
		M.~Laine and A.~Vuorinen,
		\href{https://doi.org/10.1007/978-3-319-31933-9}{Lect. Notes Phys. \textbf{925}, pp.1-281 (2016)}
		Springer, 2016,
		[arXiv:1701.01554 [hep-ph]].
		
		\bibitem{Berrehrah:2014tva}
		H.~Berrehrah, P.~B.~Gossiaux, J.~Aichelin, W.~Cassing, J.~M.~Torres-Rincon and E.~Bratkovskaya,
		\href{https://doi.org/10.1103/PhysRevC.90.051901}{Phys. Rev. C \textbf{90}, 051901 (2014)}
		[arXiv:1406.5322 [hep-ph]].
		
		\bibitem{Rapp:2008qc}
		R.~Rapp and H.~van Hees,
		\href{https://arxiv.org/abs/0803.0901}{[arXiv:0803.0901 [hep-ph]]}.
		

  \bibitem{vanHees:2007me}
H.~van Hees, M.~Mannarelli, V.~Greco and R.~Rapp,
\href{https://doi.org/10.1103/PhysRevLett.100.192301}{Phys. Rev. Lett. \textbf{100}, 192301 (2008)}
[arXiv:0709.2884 [hep-ph]].
		
		\bibitem{Riek:2010fk}
		F.~Riek and R.~Rapp,
		\href{https://doi.org/10.1103/PhysRevC.82.035201}{Phys. Rev. C \textbf{82}, 035201 (2010)}
		[arXiv:1005.0769 [hep-ph]].
		
		


  \bibitem{Liu:2016ysz}
S.~Y.~F.~Liu and R.~Rapp,
\href{https://doi.org/10.1140/epja/s10050-020-00024-z}{Eur. Phys. J. A \textbf{56}, no.2, 44 (2020)}
[arXiv:1612.09138 [nucl-th]].
		
		


  \bibitem{Scardina:2017ipo}
F.~Scardina, S.~K.~Das, V.~Minissale, S.~Plumari and V.~Greco,
\href{https://doi.org/10.1103/PhysRevC.96.044905}{Phys. Rev. C \textbf{96}, no.4, 044905 (2017)}
[arXiv:1707.05452 [nucl-th]].
		
		\bibitem{Banerjee:2011ra}
		D.~Banerjee, S.~Datta, R.~Gavai and P.~Majumdar,
		\href{https://doi.org/10.1103/PhysRevD.85.014510}{Phys. Rev. D \textbf{85}, 014510 (2012)}
		doi:10.1103/PhysRevD.85.014510
		[arXiv:1109.5738 [hep-lat]].
		


  \bibitem{Ghosh:2011bw}
S.~Ghosh, S.~K.~Das, S.~Sarkar and J.~e.~Alam,
\href{https://doi.org/10.1103/PhysRevD.84.011503}{Phys. Rev. D \textbf{84}, 011503 (2011)}
[arXiv:1104.0163 [nucl-th]].
		
		\bibitem{Tolos:2013kva}
		L.~Tolos and J.~M.~Torres-Rincon,
		\href{https://doi.org/10.1103/PhysRevD.88.074019}{Phys. Rev. D \textbf{88}, 074019 (2013)}
		doi:10.1103/PhysRevD.88.074019
		[arXiv:1306.5426 [hep-ph]].
		
		\bibitem{He:2011yi}
		M.~He, R.~J.~Fries and R.~Rapp,
		\href{https://doi.org/10.1016/j.physletb.2011.06.019}{Phys. Lett. B \textbf{701}, 445-450 (2011)}
		[arXiv:1103.6279 [nucl-th]].
		


  \bibitem{Torres-Rincon:2021yga}
J.~M.~Torres-Rincon, G.~Monta\~na, \`A.~Ramos and L.~Tolos,
\href{https://doi.org/10.1103/PhysRevC.105.025203}{Phys. Rev. C \textbf{105}, no.2, 025203 (2022)}
[arXiv:2106.01156 [hep-ph]].
		


  \bibitem{Abreu:2011ic}
L.~M.~Abreu, D.~Cabrera, F.~J.~Llanes-Estrada and J.~M.~Torres-Rincon,
\href{https://doi.org/10.1016/j.aop.2011.06.006}{Annals Phys. \textbf{326}, 2737-2772 (2011)}
[arXiv:1104.3815 [hep-ph]].
		


  \bibitem{Das:2011vba}
S.~K.~Das, S.~Ghosh, S.~Sarkar and J.~e.~Alam,
\href{https://doi.org/10.1103/PhysRevD.85.074017}{Phys. Rev. D \textbf{85}, 074017 (2012)}
[arXiv:1109.3359 [hep-ph]].
		

  \bibitem{Abreu:2012et}
L.~M.~Abreu, D.~Cabrera and J.~M.~Torres-Rincon,
\href{https://doi.org/10.1103/PhysRevD.87.034019}{Phys. Rev. D \textbf{87}, no.3, 034019 (2013)}
[arXiv:1211.1331 [hep-ph]].
		


  \bibitem{Ghosh:2014oia}
S.~Ghosh, S.~K.~Das, V.~Greco, S.~Sarkar and J.~e.~Alam,
\href{https://doi.org/10.1103/PhysRevD.90.054018}{Phys. Rev. D \textbf{90}, no.5, 054018 (2014)}
[arXiv:1407.5069 [nucl-th]].
		

\bibitem{Tolos:2016slr}
L.~Tolos, J.~M.~Torres-Rincon and S.~K.~Das,
\href{https://doi.org/10.1103/PhysRevD.94.034018}{Phys. Rev. D \textbf{94}, no.3, 034018 (2016)}
[arXiv:1601.03743 [hep-ph]].

  
		\bibitem{Csernai:2006zz}
		L.~P.~Csernai, J.~I.~Kapusta and L.~D.~McLerran,
		\href{https://doi.org/10.1103/PhysRevLett.97.152303}{Phys. Rev. Lett. \textbf{97}, 152303 (2006)}
		[arXiv:nucl-th/0604032 [nucl-th]].
		

\bibitem{Abhishek:2017pkp}
A.~Abhishek, H.~Mishra and S.~Ghosh,
\href{https://doi.org/10.1103/PhysRevD.97.014005}{Phys. Rev. D \textbf{97}, no.1, 014005 (2018)}
[arXiv:1709.08013 [hep-ph]].
  

\bibitem{Singha:2017jmq}
P.~Singha, A.~Abhishek, G.~Kadam, S.~Ghosh and H.~Mishra,
\href{https://doi.org/10.1088/1361-6471/aaf256}{J. Phys. G \textbf{46}, no.1, 015201 (2019)}
[arXiv:1705.03084 [nucl-th]].
  


  \bibitem{Sasaki:2008um}
C.~Sasaki and K.~Redlich,
\href{https://doi.org/10.1016/j.nuclphysa.2009.11.005}{Nucl. Phys. A \textbf{832}, 62-75 (2010)}
[arXiv:0811.4708 [hep-ph]].
		

\bibitem{Deb:2016myz}
P.~Deb, G.~P.~Kadam and H.~Mishra,
\href{https://doi.org/10.1103/PhysRevD.94.094002}{Phys. Rev. D \textbf{94}, no.9, 094002 (2016)}
[arXiv:1603.01952 [hep-ph]].
  
		\bibitem{Chakraborty:2010fr}
		P.~Chakraborty and J.~I.~Kapusta,
		\href{https://doi.org/10.1103/PhysRevC.83.014906}{Phys. Rev. C \textbf{83}, 014906 (2011)}
		[arXiv:1006.0257 [nucl-th]].
		
		\bibitem{Moore:2004tg}
		G.~D.~Moore and D.~Teaney,
		\href{https://doi.org/10.1103/PhysRevC.71.064904}{Phys. Rev. C \textbf{71}, 064904 (2005)}
		[arXiv:hep-ph/0412346 [hep-ph]].
		
		
		
		
		
		
		
		
		
		
		
		
		
		
		
		
		
		
		
		
		
		
		
		
		
		
	\end{thebibliography}
\end{document}